\documentclass[aps,twocolumn,showpacs,amssymb,pra]{revtex4-1}
\usepackage{amsmath}
\usepackage{graphicx}
\usepackage{dcolumn}
\usepackage{bbm}
\usepackage{color}

\newcommand{\Tr}{\mbox{Tr}}

\newcommand{\zero}{_{(0)}}

\DeclareMathOperator{\im}{Im}

\usepackage[normalem]{ulem} 

\begin{document}
\title{Characterization of Mott-insulating and superfluid phases\\
in the one-dimensional Bose--Hubbard model}

\author{Satoshi Ejima}
\author{Holger Fehske}
\affiliation{Institut f{\"u}r Physik,
             Ernst-Moritz--Arndt Universit{\"a}t Greifswald,
             D-17489 Greifswald,
             Germany}

\author{Florian Gebhard}
\author{Kevin zu M\"unster}
\affiliation{Fachbereich Physik, Philipps Universit\"at Marburg,
D-35032 Marburg, Germany}

\author{Michael Knap}
\author{Enrico Arrigoni}
\author{Wolfgang von der Linden}
\affiliation{Institute of Theoretical and Computational Physics, 
Graz University of Technology, A-8010 Graz, Austria}
\begin{abstract}
We use strong-coupling perturbation theory, the 
variational cluster approach (VCA),
and the dynamical density-matrix renormalization group (DDMRG) method to 
investigate static and dynamical properties of the one-dimensional 
Bose--Hubbard model in both the Mott-insulating and superfluid phases. From
the von Neumann entanglement entropy we determine
the central charge and the transition points for the first two Mott lobes.
Our DMRG results for the ground-state energy, 
momentum distribution function, boson correlation function decay, 
Mott gap, and single particle-spectral function are 
reproduced very well by the strong-coupling expansion to fifth order,
and by VCA with clusters up to 12 sites
as long as the ratio between the hopping amplitude and 
on-site repulsion, $t/U$, is smaller than 0.15 and 0.25, respectively. 
In addition, in the superfluid phase VCA captures 
well the ground-state energy
and the sound velocity of the linear phonon modes.
This comparison provides an authoritative estimate
for the range of applicability of these methods. 
In strong-coupling theory for the Mott phase, the dynamical structure factor 
is obtained from the solution of an effective single-particle problem
with an attractive potential. The resulting
resonances show up as 
double-peak structure close to the Brillouin zone boundary.
These high-energy features also appear in the superfluid phase
which is characterized by a pronounced phonon mode
at small momenta and energies, as predicted by Bogoliubov and field theory.
In one dimension, there are no traces of an amplitude mode
in the dynamical single-particle and two-particle correlation functions.
\end{abstract}

\maketitle

\section{Introduction}
\label{sec:I}
The ability to place ultracold bosonic atoms in optical lattices 
offered new prospects in the study of quantum many-particle 
systems~\cite{GMEHB02,GMHB02}, mainly because,
in contrast to solid-state realizations,
the properties of the system can be manipulated 
in a very controlled way by tuning the particle density, the 
lattice depth, the trapping potential and the interactions between 
the particles~\cite{KSDZ04,BDZ08}. 
Likewise, the spatial dimension and coordination number of 
the optical lattice, the degree of disorder, or the coupling strength to 
external fields might be changed~\cite{GHM02,LCBALS07}. 
Hence, in these experiments, specific lattice Hamiltonians 
can be engineered and analyzed, including 
quantum phase transitions between gapped and itinerant phases. 
A prominent example is
the transition between Mott insulating (MI) and superfluid
(SF) phases which results from the competition between the particles'
kinetic energy and their mutual on-site repulsion. In this way, 
subtle quantum correlation effects become observable 
on a macroscopic scale. 

The Bose--Hubbard Hamiltonian captures 
the essential physics of interacting bosons 
in optical lattices~\cite{JBCGZ98}. The ground-state 
phase diagram of this model in two and three dimensions
has been determined  by analytical, perturbative
methods~\cite{FWGF89,THHE09,FKKKT09} and numerical, 
quantum Monte-Carlo techniques~\cite{KT91,BRSRMDT02,PEH09,CPS07,KZKT08}. 
The one-dimensional (1D) case, which can be realized 
experimentally~\cite{SMSKE04}, is also accessible 
by QMC~\cite{BSZ90}, and particularly rewarding to study because
the physics in 1D normally is rather peculiar~\cite{Gi04}.  

On a linear chain with $L$ sites and periodic boundary conditions (PBC)  
the Bose--Hubbard Hamiltonian reads
\begin{eqnarray}
\hat H&=& t \hat{T} + U \hat{D} \nonumber \; ,\\
\hat{T} &=& -\sum_{j=1}^L( \hat{b}_j^\dagger \hat{b}_{j+1}^{\phantom{\dagger}}
          +\hat{b}_{j+1}^{\dagger}\hat{b}_{j}^{\phantom{\dagger}} )
  \label{hamil} \; ,\\
\hat{D} &=& \frac{1}{2}\sum_{j=1}^{L} \hat{n}_j(\hat{n}_j-1)\; .
\nonumber
\end{eqnarray}
Here, $\hat{b}_j^{\dagger}$, $\hat{b}_j^{\phantom{\dagger}}$,
and $\hat{n}_j=\hat{b}_j^\dagger \hat{b}_j^{\phantom{\dagger}}$ 
are the boson creation, annihilation and particle number 
operators on site $j$. 

The grand-canonical Hamiltonian is given by $\hat{K}=\hat{H}-\mu \hat{N}$ 
where $\mu$ is the thermodynamic
chemical potential and $\hat{N}=\sum_j\hat{n}_j$ counts the total number
of particles. For $N$~particles (atoms) in the system, 
the (global) filling factor is $\rho=N/L$.  

In Eq.~\eqref{hamil}, the hopping of the bosons between neighboring sites 
is characterized by the tunneling amplitude $t$, while 
$U$~is the on-site interaction which we choose to be repulsive, 
$U>0$; recently, N\"agerl et al.\ investigated an unstable crystal 
of bosons with $U<0$~\cite{Naegerl-2012}.
Accordingly, the physics of the Bose--Hubbard model is governed by the ratio 
between kinetic energy and interaction energy, $x=t/U$. If,
for given chemical potential $\mu$, $x$ is larger 
than a critical value the bosons are ``superfluid''. Below~$x_{\rm c}$,
the system becomes Mott insulating, characterized by an integer filling
factor $\rho$. In experiments,  
$x$~can be varied over several orders of magnitude, by modifying 
the depth of the lattice through quantum optical techniques
whereby SF and MI phases can be realized. From a theoretical point of view,
the calculation of the boundaries between the SF and MI phases 
in the $(\mu,U)$ ground-state phase diagram is particularly demanding 
because quantum phase transitions in one dimension 
often are of Kosterlitz--Thouless type~\cite{Gi04}
with exponentially small Mott gaps in the vicinity of the transition.

\begin{figure}[htbp]
\begin{center}
\includegraphics[width=0.7\linewidth,clip]{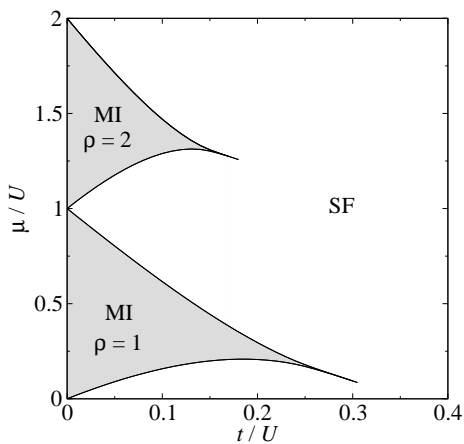}
\end{center}
\caption{Phase diagram of the 1D Bose--Hubbard model 
showing superfluid (SF) and Mott insulating (MI) regions
as a function of the chemical potential~$\mu/U$ 
and the electron transfer amplitude~$t/U$.
The boundaries delimiting the first two Mott lobes
were determined by DMRG, using system sizes up to
$L=128$ and OBC~\cite{EFG11}, see text. 
\label{fig:pd}}
\end{figure}
 
The numerical density-matrix renormalization group (DMRG) 
method~\cite{Wh92,Wh93} is well suited to address 
one-dimensional interacting particle systems~\cite{KM98,KWM00,KSDZ04}.
In fact, the $(\mu,U)$ ground-state phase diagram of the 1D Bose--Hubbard model
has been obtained fairly accurately
using this technique~\cite{EFG11}, see Fig.~\ref{fig:pd}.
Since multiple occupancies pose serious
technical problems, the maximal boson number per site in DMRG
is constrained to be five. Note that 
DMRG naturally works at fixed $N,L <\infty$, i.e.,
in the canonical ensemble.
This leads to the definition of two chemical potentials for finite systems,
$\pm\mu^{\pm}(L)=E_0(L, N\pm 1)-E_0(L,N)$~\cite{KWM00} 
where $E_0(L,N)$ denotes the ground-state energy.
In the MI state we have a finite gap, 
$\Delta=\mu^+(L\to\infty)-\mu^-(L\to\infty)>0$, whereas
the chemical potential is continuous
in the SF phase, $\mu=\mu^+(L\to\infty)=\mu^-(L\to\infty)$.

In one dimension, the delocalized SF state is not macroscopically occupied 
but rather characterized by an algebraic divergence of the momentum
distribution~\cite{MC03,Gi04}. The localized MI state is incompressible,
as usual, and characterized by an integer particle density and a gap in the 
single-particle spectrum~\cite{FWGF89}. The regions 
in the $(\mu,U)$ phase diagram where the 
density $\rho$ is pinned to integer values are termed Mott lobes.
Their special shape is conditioned by the strong phase fluctuations
existing in a 1D system. Close to the boundaries of the Mott lobes,
the Mott gap is exponentially small. The precise 
position of the Mott dips can be 
obtained from the Tomonaga--Luttinger parameter~\cite{KWM00,EFG11}.

A detailed theoretical understanding of the Bose--Hubbard model 
requires the calculations of (dynamical) correlation functions which
poses a hard problem for which no exact solution exists.
Recall that the 1D Bose--Hubbard model at $U< \infty$
(soft-core bosons) is not integrable.
Consequently, a large variety of approximative approaches 
were suggested and elaborated for the Bose--Hubbard model
and its variants during the last two decades; for a recent 
review see~Ref.~[\onlinecite{CCGOR11}]. 

In the SF phase, (weakly) interacting bosons at low energies
are well described as a Tomonaga--Luttinger liquid~\cite{FWGF89,Gi92}. 
However, close to the SF to MI transition, the precise character of 
the spectrum is still under debate.
This particularly concerns the question whether or not a second, gapped mode
besides the standard sound mode, as obtained from 
mean-field theory~\cite{HABB07}, can be seen
in the single-particle spectral function or in the dynamical structure 
factor.

In the MI phase, strong-coupling expansions in $x=t/U$ give reliable 
analytical results. The ground-state energy of all Mott lobes
was determined to second order by Freericks and Monien~\cite{FM96},
and was improved up to order $x^{14}$ for the lowest Mott lobe, $\rho=1$, 
by Damski and Zakrzewski~\cite{DZ06}.
They also provided the series expansion for the local particle-density
fluctuations to order $x^{13}$, a high-order series expansion
for the single-particle density matrix $P(r)={\cal O}(x^r)$ 
for $r=1,2,3$, and gave the corresponding expressions
for the ground-state density-density correlation function 
$D(r)-1={\cal O}(x^{2r})$ for $r=1,2,3$; for results for $r\leq 6$ and
$r\leq 10$, respectively, see Ref.~\cite{THHE09}.
The Fourier transformation of $P(r)$ provides the 
momentum distribution $n(k)$. The result for $n(k)$
to third order in~$x$ was re-derived by Freericks 
et al.~\cite{FKKKT09} using a different method.

In contrast to higher dimensions, $d\geq 2$, 
the convergence of the strong-coupling expansion series 
in 1D is rather questionable.
These problems become apparent, e.g., in the calculation 
of the critical value $x_{\rm c}$ for the transition between the
Mott insulator and the superfluid phase. For example, the 
series expansion for the superfluid susceptibility constructed by 
Eckardt et al.~\cite{THHE09} determines $x_{\rm c}$ very accurately
in $d\geq 2$ but fails for $d=1$ where a reentrant superfluid phase
is predicted~\cite{THHE09,KM98}. 

High-order expansions are also possible for the single-particle 
gap~\cite{EM99b}. The Mott transition in one dimension
is of Kosterlitz--Thouless (KT) 
type so that the gap becomes exponentially small 
close to the transition, 
which cannot be reproduced easily within a third-order 
strong-coupling expansion~\cite{FM96}.
In order to obtain a good approximation of the critical value for
the transition, Elstner and Monien~\cite{EM99} proposed
a scaling analysis for the gap. Based on this idea, 
Freericks et al.~\cite{FKKKT09} used a (6,7)--Pad\'e approximant 
for the square of the
logarithm of the single-particle gap to find $x_{\rm c}\approx 0.300(1)$
for $\rho=1$, in good agreement with the DMRG value; for another scheme,
see Heil and von der Linden~\cite{CH-WvdL}.

In the present paper, we first refine and extend the perturbative 
strong-coupling approach in order to analyze 
the single-particle spectral function and the dynamical structure factor.  
For the latter quantity, we obtain higher-order corrections from 
the corresponding Green's function. Secondly, in order to relax 
the strong-coupling condition, we employ the 
variational cluster approach (VCA) that is applicable in both
the Mott insulating and the superfluid 
phase~\cite{KD06,KAL10a,KAL11,AKL11}. For the calculation
of spectral properties in the SF phase, the VCA can be reformulated in terms
of a pseudo-particle approach, whereby single-particle excitations within 
a cluster are approximately mapped onto particle-like 
excitations~\cite{KAL11}, or in terms of the self-energy 
functional approach~\cite{AKL11,Pot03a}. Thirdly, we perform large-scale DMRG
calculations: (i) to access the whole parameter space of the 
Bose--Hubbard model and (ii) to benchmark the reliability of the 
used analytical strong-coupling and numerical VCA techniques.    
While in the past DMRG has been successfully applied 
to investigate the ground-state properties of the Bose--Hubbard 
model~\cite{KM98,KWM00,KSDZ04}, DMRG results for dynamical 
properties at zero temperatures are rare (in contrast 
to fermionic systems), but highly desirable because superfluids 
in optical lattices can be studied by momentum-resolved Bragg 
spectroscopy~\cite{clement_exploring_2009,fabbri_excitations_2009,EGKPLPS09,%
du_bragg_2010}.
 
The outline of this paper is as follows. In Sect.~\ref{sec:PA} 
we describe perturbative approaches to the Bose--Hubbard model, 
and present a detailed derivation of the strong-coupling 
results for static and dynamical quantities. Sects.~\ref{sec:VCA} 
and~\ref{sec:DMRG} sketch the specifics of the VCA and DMRG, 
respectively, when applied to the Bose--Hubbard model. Sect.~\ref{sec:results}
contains our main results. In particular, we discuss how the 
von Neumann entanglement entropy can be calculated from DMRG and 
how it can be used to determine the KT transition point in 
the Mott lobes. Next, we determine the ground-state energy,
the boson correlation function, and the momentum distribution function. 
Lastly, we analyze the photoemission spectra and dynamical structure
factors. In all cases, 
we compare analytical and numerical results. Finally, 
Sect.~\ref{sec:S} summarizes our findings.

\section{Perturbative Approaches}
\label{sec:PA}

\subsection{Weak-coupling limit}
\label{subsec:WC}

For weak interactions, we use the perturbative results obtained 
by Bogoliubov~\cite{Bo47} 
(see Fetter and Walecka, chap.~35~\cite{FW71}),
for a weakly interacting Bose gas with contact interaction
and density $\rho=N/L$. From the text-book formulae
we find for the 1D Bose--Hubbard Hamiltonian at $\rho=1$ 
\begin{eqnarray}
\epsilon(k)&=& -2t(\cos(k)-1) \; , \label{be}\\
E(k)&=& \sqrt{\epsilon(k)(\epsilon(k)+2U)}\; ,\label{bE}\\
\frac{N_0}{L}&=& 1- \frac{1}{2L}\sum_{k\neq 0} 
\left( \frac{\epsilon(k)+U}{E(k)}-1\right) 
\label{ooops}\;,
\end{eqnarray}
where $\epsilon(k)$ is the bare dispersion of~(\ref{hamil}) shifted by $2t$,
$E(k)$ is the dispersion of the Bogoliubov quasiparticles,
and $N_0$ is the number of particles in the condensate.
Here, $k=2\pi m_k/L, m_k=0,1,\ldots, L-1$ are the crystal momenta for PBC.
The Bogoliubov ground-state energy reads
\begin{equation}
\frac{E_0^{\rm B}(U)}{L}= -2t +\frac{U}{2} + \frac{1}{2L}\sum_{k\neq 0}
\left(E(k)-\epsilon(k)-U \right) \; .
\label{bgs}
\end{equation}

Two problems with the Bogoliubov theory become apparent when we
consider some limits. First, we address the limit $k\to 0$ for $E(k)$,
$E(k\to 0)\sim k$.
Therefore, in the thermodynamic limit,
the integral in Eq.~(\ref{ooops}) is logarithmically divergent,
and $N_0=0$ in 1D results, in agreement with field theory~\cite{Gi04}.
This, however, invalidates the starting point of the Bogoliubov approximation.
Second, we cannot apply the theory for large $U/t$ because
$E(k,U\gg t)\approx \sqrt{8Ut}|\sin(k/2)|$ so that
$(E_0^{\rm B}(U\gg t)/L)\sim \sqrt{Ut}$ for large $U/t$, 
in contrast with the exact limit, $\lim_{U\to\infty}E_0(U)=0$. 

The analytical result for the ground-state energy in Bogoliubov theory
is found, e.g.~using {\sc Mathematica}~\cite{Mathem}, as
\begin{equation}
\frac{E_0^{\rm B}(U)}{L} = -3t+\frac{\sqrt{2Ut}}{\pi}
+\frac{U+2t}{\pi}\arccos\left(\sqrt{\frac{U}{U+2t}}\right) \; .
\label{fullformula}
\end{equation}
The small-$U$ expansion is
\begin{equation}
\frac{E_0^{\rm B}(U\ll t)}{Lt} = -2+\frac{U}{2t}
-\frac{\sqrt{2}(U/t)^{3/2}}{3\pi} \; .
\label{sue}
\end{equation}
Corrections are of the order $(U/t)^{5/2}$ which is formally
beyond the validity
of the Bogoliubov expansion which ignores terms of the order $(U/t)^2$.

\subsection{Strong-coupling limit}
\label{subsec:SC}

\subsubsection{Harris--Lange transformation}
\label{subsubsec:HLT}
For the bosonic Hubbard model, an $x=t/U$ strong-coupling expansion 
easily permits the calculation of the ground state for $x\to 0$,
\begin{equation}
|\phi_0\rangle= \frac{1}{(\rho !)^{L/2}}
\prod_{i} \left(\hat{b}_i^{\dagger}\right)^{\rho} |{\rm vac}\rangle\;,
\label{defphi0}
\end{equation}
because it is non-degenerate for the Mott lobe with integer filling 
$\rho=N/L$.  
Likewise, 
the energy levels of a single-hole excitation, $E_{\rm h}(k)$, 
and of a single-particle excitation, $E_{\rm p}(k)$, 
can be determined to high order in~$x$ because the 
perturbation theory for these energy levels also starts 
from non-degenerate states, e.g., for $\rho=1$,
\begin{eqnarray}
|\phi_{\rm h}(k)\rangle&=& 
\sqrt{\frac{1}{L}}
\sum_{l=1}^L e^{-{\rm i}kl} \hat{b}_l |\phi_0\rangle\; ,
\label{defphi-hole}\\
|\phi_{\rm p}(k)\rangle&=& \sqrt{\frac{1}{L}}\sqrt{\frac{1}{2!}}
\sum_{l=1}^L e^{{\rm i}kl} \hat{b}_l^{\dagger} |\phi_0\rangle\; .
\label{defphi-particle}
\end{eqnarray} 

When we employ the unitary Harris--Lange transformation~\cite{HL67},
the strong-coupling Hamiltonian of the Bose--Hubbard model 
can be derived in a systematic way,
\begin{eqnarray}
\hat{h}&=&e^{\hat{S}} \hat{H} e^{-\hat{S}} 
= U \hat{D} +t \sum_{r=0}^{\infty}x^r \hat{h}_r \; ,\\
\hat{S}&=&-\hat{S}^{\dagger}=
\sum_{r=1}^{\infty}x^r \hat{S}_r \; .
\end{eqnarray}
In practice, a finite order in the expansion of $\hat{S}$
is kept. When we retain $\hat{S}_r$ for $1\leq r\leq n$, we 
denote this the `$n$th-order approximation'.
In $n$th order we thus keep $(n-1)$ terms in the expansion for $\hat{h}$
whose terms obey $[\hat{h}_r,\hat{D}]_{-}=0$ for $0\leq r\leq n-1$.
To order $(n-1)$, the number of double occupancies is conserved by $\hat{h}$.
This defines the construction principle for 
the operators $\hat{S}_n$.

The leading order terms for $\hat{S}_r$ and $\hat{h}_r$ are given by
\begin{eqnarray}
\hat{S}_1&=&\sum_{D_1,D_2} 
\frac{\hat{P}_{D_1} \hat{T} \hat{P}_{D_2}}{D_1-D_2}\; ,
\\[6pt]
\hat{S}_2&=&\sum_{D_1,D_2} 
\frac{-\hat{P}_{D_1} \hat{T} \hat{P}_{D_1} \hat{T} \hat{P}_{D_2}
+\hat{P}_{D_1} \hat{T} \hat{P}_{D_2} \hat{T} \hat{P}_{D_2}}{(D_1-D_2)^2}
\\[3pt]
&& +\!\! \sum_{D_1,D_2,D_3}\!\!
\frac{\hat{P}_{D_1} \hat{T} \hat{P}_{D_3} \hat{T} \hat{P}_{D_2}}{2(D_1-D_2)}
\frac{[D_1-D_3+D_2-D_3]}{(D_1-D_3)(D_2-D_3)}
\nonumber 
\; ,
\\[6pt]
\hat{h}_0&=& \sum_{D}\hat{P}_D \hat{T} \hat{P}_{D} \; ,
\label{hath0}
\\[6pt]
\hat{h}_1&=& \sum_{D_1,D_2}
\frac{\hat{P}_{D_1} \hat{T} \hat{P}_{D_2} \hat{T} \hat{P}_{D_1}}{D_1-D_2}\; ,
\end{eqnarray}
where $\hat{P}_D$ is the projection operator onto the subspace
of eigenstate with $D$ interactions, $\hat{D}=\sum_{D=0}^{\infty} D\hat{P}_D$.
In the above sums it is implicitly understood that all indices $D_i\geq 0$
are mutually different.
A compact formula for the recursive generation of higher orders
can be found in Ref.~\cite{Do94}.
In our analysis, we use a computer program to generate orders
$r\geq 2$ in $\hat{S}_r$ and $\hat{h}_r$~\cite{MueMaster12}.

For the exact ground state of the Mott insulator we have
\begin{equation}
\hat{H}|\psi_0\rangle = E_0 |\psi_0\rangle \; ,
\end{equation}
where $E_0$ is the exact ground-state energy.
Within the strong-coupling expansion we then find
\begin{equation}
|\psi_0\rangle = e^{\hat{S}}|\phi_0\rangle\quad ; \quad
\hat{h}|\phi_0\rangle = E_0 |\phi_0\rangle \; ,
\end{equation}
where $|\phi_0\rangle$ is the ground state of $\hat{h}_{-1}=\hat{D}$,
see Eq.~(\ref{defphi0}).

Since the Harris--Lange transformation is unitary,
operators and ground-state expectation values 
translate according to 
\begin{eqnarray}
|\psi_0\rangle &\mapsto&  |\phi_0\rangle \;, \nonumber\\
\hat{H} &\mapsto&  \hat{h} \; , \\
\hat{A} &\mapsto&  \widetilde{A}=e^{\hat{S}} \hat{A} e^{-\hat{S}}  \nonumber \; . 
\end{eqnarray}
The series expansion for $\hat{S}$ to $n$th order contains
$n$~powers of the kinetic energy operator~$\hat{T}$.
Therefore, local operators $\hat{A}_i$ translate into cluster operators
which involve the sites $l$ with $|l-i|\leq n$. 
The range of $\hat{h}$ scales accordingly:
the strong-coupling theory generates a cluster expansion.

\subsubsection{Static quantities}
\label{subsubsec:SQ}
For fixed momentum~$k$, the exact eigenstates of $\hat{h}$
with one extra particle or one hole
in $|\phi_0\rangle$, Eq.~(\ref{defphi0}), 
are given by the hole and particle states defined in 
Eqs.~(\ref{defphi-hole}), (\ref{defphi-particle}).
In this sector, we thus obtain the ground-state energy and the
single-particle excitation energies from
\begin{eqnarray}
E_0 &= & \langle \phi_0 |\hat{h}|\phi_0\rangle \; ,\\[3pt]
E_{\rm p}(k) &=& \langle \phi_{\rm p}(k) 
|\hat{h}|\phi_{\rm p}(k)\rangle -E_0
\; ,\\[3pt]
E_{\rm h}(k) &=& \langle \phi_{\rm h}(k) |\hat{h}|\phi_{\rm h}(k)\rangle -E_0
\; .
\end{eqnarray}
Up to and including 6th order in $x$, 
we obtain for the ground-state energy per site
\begin{equation}
\frac{E_0^{[6]}}{4UL}=-x^2+x^4+\frac{68}{9}x^6+{\cal O}(x^8)\;,
\end{equation} in agreement with Ref.~\cite{DZ06}.

The single-hole and single-particle excitations energies are
\begin{eqnarray}
\frac{E_{\rm h}(k)}{t} &=& 8x -\frac{512}{3}x^{5} \label{Ehole} \\
&&+\left(-2+12x^{2}-\frac{224}{3}x^{4}\right)\cos (k) \nonumber\\
&&+\left(-4x+64x^{3}-\frac{1436}{3}x^{5}\right)\cos (2k)\nonumber 
\\
&&+\left(-12x^{2}+276x^{4}\right)\cos (3k)  \nonumber\\
&&+\left(-44x^{3}+1296x^{5}\right)\cos (4k) \nonumber\\
&&-180x^{4}\cos (5k) -792 x^{5}\cos (6k) + {\cal O}\left({x^{6}}\right) \; ,
\nonumber
\end{eqnarray}
and
\begin{eqnarray}
\frac{E_{\rm p}(k)}{t} &=& 
\frac{1}{x}+5x-\frac{513}{20}x^{3}-\frac{80139}{200}x^{5}\label{Eparticle}\\
&&+\left(-4+18x^{2}-\frac{137}{150}x^{4}\right)\cos (k) \nonumber\\
&&+\left(-4x+64x^{3}-\frac{426161}{1500}x^{5}\right)\cos (2k) \nonumber 
\\
&&+\left(-12x^{2}+276x^{4}\right)\cos (3k)  \nonumber\\
&&+\left(-44x^{3}+1296x^{5}\right)\cos (4k) \nonumber\\
&&-180x^{4}\cos (5k) -792x^{5}\cos (6k) 
+ {\cal O}\left({x^{6}}\right) \nonumber \; .
\end{eqnarray}
The single-particle gap is calculated from $\Delta=E_{\rm p}(0)+E_{\rm h}(0)$
which results in
\begin{equation}
\frac{\Delta}{U}=1-6x+5x^{2}+6x^{3}+\frac{287}{20}x^{4}
+\frac{5821}{50}x^{5} -\frac{602243}{1000}x^{6} 
+ \ldots \; ,
\end{equation}
in agreement with Ref.~\cite{EM99b}.

\subsubsection{Single-particle spectral functions}
\label{subsubsec:SPSF}
The single-particle spectral functions are obtained from 
\begin{eqnarray}\label{A_plus}
A^+(k,\omega)&=&\sum_n \left| \langle\phi_n| 
\hat{b}^{\dagger}(k) |\phi_0\rangle
\right|^2 \delta\left(\omega-\omega_n^+\right) \; ,\\
A^-(k,\omega)&=&\sum_n \left| \langle\phi_n| 
\hat{b}^{\vphantom{\dagger}}(k) |\phi_0\rangle
\right|^2 \delta\left(\omega+\omega_n^-\right) \; ,
\label{A_minus}
\end{eqnarray}
where $\omega_n^{\pm}= E_n-E_0$ is the excitation energy
of the exact eigenstates $|\phi_n\rangle$ of $\hat{h}$
with $N=\rho L \pm 1$ bosons, measured from the ground-state energy, and
\begin{equation}
\hat{b}^{\vphantom{\dagger}}(k) =\sqrt{\frac{1}{L}} 
\sum_{l=1}^L e^{-{\rm i}kl} \hat{b}_l^{\vphantom{\dagger}} 
\; , \;
\hat{b}^{\dagger}(k) =\sqrt{\frac{1}{L}} 
\sum_{l=1}^L e^{{\rm i}kl} \hat{b}_l^{\dagger} 
\end{equation}
for PBC. Obviously, 
the single-particle gap $\Delta$ is obtained from 
$\Delta={\rm Min}_n(\omega_n^{+})-{\rm Max}_n(-\omega_n^{-})$.

For the calculation of the spectral function, we need the weight
factors
\begin{eqnarray}
w_{\rm p}(k) 
= \left| \langle \phi_{\rm p}(k) | k_{+} \rangle \right|^{2}
\; &,&  \; 
| k_{+}\rangle = \widetilde{b}_k^{\dagger}  | \phi_0 \rangle\; , 
\label{k+}
\\
w_{\rm h}(k) = \left| 
\langle \phi_{\rm h}(k)| k_{-}\rangle \right|^{2}
\; &,& \;
| k_{-}\rangle = \widetilde{b}_k^{\vphantom{\dagger}}  | \phi_0 \rangle \; .
\label{k-}
\end{eqnarray}
Up to and including third order in $x$ we find
\begin{eqnarray}
w_{\rm h}(k)  &=& \Bigl[
1-4x^{2} +\left(4x-20x^{3}\right)\cos(k) \nonumber \\
&&  +14x^{2}\cos(2k) +60x^3 \cos(3k)
\Bigr]^{2} \; ,
\label{whole}
\\
w_{\rm p}(k)  &=& 2 \Bigl[
1-\frac{7}{4} x^2 +\left(2x-\frac{15}{4} x^{3}\right)\cos(k) 
\nonumber \\
&&  +8x^{2}\cos(2k) +18x^{3} \cos(3k)
\Bigr]^{2}
\; .
\label{wparticle}
\end{eqnarray}
The weights $w_{\rm p,h}(k)$ are those of the lower and upper Hubbard bands
which are energetically closest to the single-particle gap 
and separated by~$U$ in the atomic limit. 

In higher orders of the strong-coupling expansion,
secondary Hubbard bands appear in the single-particle 
spectral function~\cite{HL67,Nishimoto2004,KAL10a,KAL10b}.
This can most easily be seen from
the weights which express the overlap
of the exact excited eigenstates of $\hat{h}$ with
the states $|k_{\pm}\rangle$~see Eqs.~(\ref{k+}),~(\ref{k-}).
With an amplitude of the order $x^{2}$,
the state $|k_{-}\rangle$ contains a component
with two neighboring holes and one doubly occupied site in a row.
This component is {\sl not\/} in the original subspace with
$D=0$ and contributes to the upper Hubbard band with weight $x^{4}$.
Therefore, for the weight of the lower Hubbard band we have
\begin{eqnarray}
w_{\rm LHB}(k)&=&w_{\rm h}(k)+{\cal O}(x^{4})\nonumber\\
&=&1+(8x-16x^3)\cos(k) +36x^{2}\cos(2k)\nonumber\\ &&+176 x^{3}\cos(3k)
+{\cal O}(x^{4})\;.
\end{eqnarray}

The state $|k_{+}\rangle$ contains configurations with
a triple occupancy and a neighboring hole to the left or right.
Their amplitude up to order $x^{2}$ is
$a_{\pm}(k)=\sqrt{6}(x/2-\exp(\pm {\rm i}k)x^{2}/3)$.
They contribute to the secondary Hubbard
band centered around $\omega=3U$ to order $x^{2}$ and $x^{3}$.
Components with two double occupancies
and a quadruple occupancies have an amplitude proportional
to order $x^{2}$ and thus contribute to the bands centered around
$\omega=2U$ and $\omega=6U$, respectively, with weights 
of the order of $x^{4}$.
Up to and including order $x^{3}$,
the secondary Hubbard band around $\omega=3U$ has the weight
\begin{equation}
w_{3U}(k)= 12 \left| \frac{x}{2} -\frac{x^{2}e^{-{\rm i}k}}{3} \right|^{2}
+ {\cal O}\left(x^{4}\right) \;.
\label{w3U}
\end{equation}
Therefore, the total weight for the upper Hubbard bands is given
by $w_{\rm UHB}(k) = w_{\rm p}(k) + w_{3U}(k)
= 1+w_{\rm LHB}(k)$, in agreement with the sum rule
\begin{equation}
\int_{-\infty}^{\infty} {\rm d}\omega [A^{+}(k,\omega)-A^{-}(k,\omega)]
=w_{\rm UHB}(k)-w_{\rm LHB}(k) =1 \; ,
\label{sr1}
\end{equation}
which follows directly from the definition of the spectral function.
Another check results from the momentum 
distribution sum rule,
\begin{equation}\label{sr2}
w_{\rm LHB}(k) =\! \int_{-\infty}^{\infty} {\rm d}\omega \!A^{-}(k,\omega)
=\langle \phi_0 | \hat{b}^{\dagger}(k)
\hat{b}^{\vphantom{\dagger}}(k)|\phi_0 \rangle = n(k) \, .
\end{equation}
Up to and including third order in $x$, our results for $n(k)$
agree with those found in Refs.~\cite{DZ06,FKKKT09}.

\subsubsection{Dynamical structure factor}
\label{subsubsec:DSF}
For the density-density correlation function we focus on $\omega>0$
so that we do not have to consider terms 
of the form $\langle \phi_0 | \widetilde{n}_{l+r}
\delta\left(\omega-(\hat{h}-E_0)\right)|\phi_0\rangle \sim \delta(\omega)$.
We define the states
\begin{equation}
|q\rangle = \left(\widetilde{n}_q-\hat{n}_q\right) |\phi_0\rangle \; , 
\label{qstatesdef}
\end{equation}
where the density operator in momentum space is given by
($q=2\pi m_q/L, m_q=0,1,\ldots,L-1$)
\begin{equation}
\hat{n}(q) = \sum_{l=1}^Le^{{\rm i}q l}\hat{n}_l 
=\sum_{k}\hat{b}^{\dagger}(k+q) \hat{b}^{\vphantom{\dagger}}(k)
= (\hat{n}(-q))^{\dagger}\; .
\end{equation}
Then, we can express the dynamical structure factor in the form
\begin{eqnarray}
S(q,\omega>0)&=& \sum_{l=1}^L e^{-{\rm i}q l}S_l(\omega)
= \langle q | \delta\left(\omega-(\hat{h}-E_0)\right) | q\rangle 
\nonumber\\
&=& \sum_{n} \left| \langle \Phi_n | q \rangle \right|^2 
\delta\left(\omega-(\hat{h}-E_0)\right) \; ,
\label{Skomeagfromexacteigenstates}
\end{eqnarray}
where $|\Phi_n\rangle$ are the exact eigenstates
of $\hat{h}$ in the sector with $N=\rho L$ bosons.

{\it Leading order contribution.} The dynamical structure factor was
calculated analytically 
within mean-field theory~\cite{HABB07},
bosonization~\cite{ICHG06,CCGOR11}, and 
lowest-order strong-coupling theory~\cite{ICHG06,TG11,EFG12}.

The strong-coupling result to leading order
is readily obtained from the exact eigenstates 
of~$\hat{h}_0$, Eq.~(\ref{hath0}), in the sector with one hole and one
double occupancy. The subspace $(N=L, D=1)$ is spanned by the 
$L(L-1)$ orthonormal states ($l\neq L$)
\begin{equation}
|q,l\rangle = \sqrt{\frac{1}{L}} \sum_{s=1}^{L} e^{{\rm i} q s} |s,l\rangle \;, 
\; |s,l\rangle= \sqrt{\frac{1}{2}}
\hat{b}_s^{\dagger}\hat{b}_{s+l}^{\vphantom{\dagger}} |\phi_0\rangle \; .
\end{equation}
The states $|q,l\rangle$ obey the effective single-particle
Schr\"o\-din\-ger equation
\begin{eqnarray}
\hat{h}_0 |q,l\rangle &=&
-(1-\delta_{l,1})(1+2e^{-{\rm i}q})|q,l-1\rangle \nonumber \\
&&-(1-\delta_{l,L-1})(1+2e^{{\rm i}q})|q,l+1\rangle \;.
\label{pairmotion}
\end{eqnarray}
As expected for a translational invariant system, the 
center-of-mass momentum $q=2\pi m_q/L$ with $m_q=0,1,\ldots,L-1$ is conserved.

The leading-order contribution to the states 
$|q\rangle=\sum_{n=1}^{\infty}x^n |q^{[n]}\rangle$ from~(\ref{qstatesdef})
is given by
\begin{equation}
|q^{[1]}\rangle = \sqrt{2}\left[ (1-e^{{\rm i }q})|q,1\rangle +
(1-e^{-{\rm i }q})|q,L-1\rangle \right] \; ,
\label{qstatesleadingorder}
\end{equation}
i.e., double occupancy and hole are nearest neighbors.

Eq.~(\ref{pairmotion}) describes a single particle
on an open chain with $L-1$ sites which reflects the fact that
hole and double occupancy cannot be on the same site,  $l\neq L$.
In contrast to the fermionic case, the hard-core constraint
is not sufficient to determine the phase shift between hole
and double occupancy because their
tunnel amplitudes differ by a factor of two.
Therefore, the scattering phase shift between double
occupancy and hole is not trivial~\cite{ICHG06,TG11,EFG12}.
This is in contrast to the mean-field 
approach~\cite{HABB07} where the bare
dispersions for hole and double occupancy 
enter Eq.~(\ref{Skomeagfromexacteigenstates}).

The normalized double-occupancy--hole eigenstates are given by
\begin{equation}
|q;k\rangle =\sqrt{\frac{2}{L}}\sum_{l=1}^{L-1} 
\sin(kl) e^{{\rm i}\phi(q) l} |q,l\rangle
\end{equation}
with $k=(\pi/L)m_k$ ($m_k=1,2,\ldots,L-1$) 
where the two-particle phase shift $\phi(q)$ follows from
\begin{equation}
\tan[\phi(q)]= \frac{2\sin(q)}{1+2\cos(q)} \; .
\end{equation}
The energies of the eigenstates $|q;k\rangle$ of $t \hat{h}_0$ 
are given by
\begin{equation}
E(q,k)= -2t\cos(k)\sqrt{5+4\cos(q)} \; .
\end{equation}
The overlap with the states in Eq.~(\ref{qstatesleadingorder})
defines the oscillator strengths in Eq.~(\ref{qstatesdef}),
\begin{eqnarray}
\langle q;k| q^{[1]}\rangle &=&\sqrt{2}
\sqrt{\frac{2}{L}}\sin(k) \Bigl[
(1-e^{{\rm i}q})e^{-{\rm i}\phi(q)} \nonumber \\
&& -(1-e^{-{\rm i}q})(-1)^{m_k}e^{-{\rm i}\phi(q)(L-1)} 
\Bigr]\; , 
\end{eqnarray}
so that, in the thermodynamic limit ($L\to\infty$), we obtain
for the weights
\begin{equation}
w(q;k)=
\left(\frac{t}{U}\right)^2 \frac{32}{L}\sin^2(k)\sin^2(q/2) \; ,
\end{equation}
where we dropped the cross terms 
because their contribution to the structure factor vanishes 
due to the fast oscillations of $(-1)^{m_k}$.

The dynamical structure factor becomes for $\omega>0$
\begin{eqnarray}
S^{[1]}(q,\omega)&=& 2 \left(\frac{4t\sin(q/2)}{U}\right)^2 
\label{finalS1}
\\
&&\times \int_{0}^{\pi} \frac{{\rm d}k}{\pi} \sin^2(k)
\delta\left(\omega-U-E(q,k)\right)\; .\nonumber 
\end{eqnarray}
Finally, for $|\omega-U|\leq 2t\sqrt{5+4\cos(q)}$ we obtain ($t\equiv 1$)
\begin{equation}
S^{[1]}(q,\omega)= \biggl(\frac{4\sin(q/2)}{U}\biggr)^2\!
\frac{\sqrt{20+16\cos(q)-(\omega-U)^2}}{2\pi(5+4\cos(q))}
\end{equation}
for the dynamical structure factor to leading 
order~\cite{ICHG06,TG11,EFG12}.

{\it Second-order and higher-order contributions.}
For the next order in the $(t/U)$-expansion we must calculate
the action of $\overline{h}_1\equiv \hat{h}_1-E_0^{[1]}$ 
on the states $|q,l\rangle$, where we use $E_0=t\sum_{n=1}^{\infty}x^nE_0^{[n]}$.
The correction to Eq.~(\ref{pairmotion}) reads
\begin{eqnarray}
\overline{h}_1|q,l\rangle
&=&13 |q,l\rangle \nonumber\\
&& -2\left[(1-\delta_{l,1})(1-\delta_{l,2})(1+e^{-2{\rm i}q})|q,l-2\rangle\right]
\nonumber \\
&& -2 \left[(1-\delta_{l,L-1})(1-\delta_{l,L-2})(1+e^{2{\rm i}q})|q,l+2\rangle\right]
\nonumber
\\
&&+ 2\delta_{l,1} \Bigl[
(-\frac{1}{4}+e^{-{\rm i}q}+e^{{\rm i}q}) |q,1\rangle
\nonumber \\
&&
\hphantom{+ 2\delta_{l,1} \Bigl[ }
+ (\frac{1}{4}+e^{-{\rm i}q}+e^{-2{\rm i}q}) |q,L-1\rangle \Bigr]
\nonumber \\
&&
+ 2\delta_{l,L-1} \Bigl[
(\frac{1}{4}+e^{{\rm i}q}+e^{2{\rm i}q}) |q,1\rangle
\nonumber \\
&&
\hphantom{+ 2\delta_{l,L-1} \Bigl[ }
+ (-\frac{1}{4}+e^{-{\rm i}q}+e^{{\rm i}q}) |q,L-1\rangle \Bigr]
\; .
\label{nextorderschroedinger}
\end{eqnarray}
The effective single-particle problem 
contains an overall energy shift $13t^2/U$,
a nearest-neighbor transfer $l\to (l+1)$ 
with amplitude $t(q)=(-t)(1+2\cos(q)+2{\rm i}\sin(q))$
as before, an additional next-nearest neighbor transfer from $l\to l+2$
with amplitude $m(q)=-2(t^2/U)(1+\cos(2q)+{\rm i}\sin(2q))$,
and a potential at the chain ends,
$V_{1,1}=V_{L-1,L-1}=(t^2/U)(4\cos(q)-1/2)$ and $V_{1,L-1}=V_{L-1,1}^*=(2t^2/U)
(1/4+\cos(q)+\cos(2q)+{\rm i}\sin(q)+{\rm i}\sin(2q))$.

Now that the potential links the two chain ends, it is computational
advantageous to treat the problem on a ring instead of a chain.
The potential is readily generalized according to 
Eq.~(\ref{nextorderschroedinger}). For a ring, the potential also
contains the terms $V_{1,L-2}$ and $V_{2,L-1}$ and their complex conjugates
so that the potential links four neighboring sites.
Moreover, the extension of $\hat{h}_0$ from a chain to a ring generates
corrections to $V_{1,L-1}$ and $V_{L-1,1}$.

The $x^2$-corrections to the states $|q\rangle$~(\ref{qstatesdef}) read
\begin{equation}
| q^{[2]}\rangle = 3\sqrt{2} 
\left[ (1-e^{2{\rm i }q})|q,2\rangle +
(1-e^{-2{\rm i }q})|q,L-2\rangle \right] \; .
\label{qstatessecondorder}
\end{equation}
As the potential $V_{a,b}$, 
the dynamical structure factor to second order involves
the four neighboring sites $l=L-2,L-1,1,2$.

The calculation of all eigenstates 
of $\hat{h}_0+\hat{h}_1$ is not feasible in the thermodynamic limit.
To calculate the dynamical structure factor we
address the corresponding Green's function
\begin{equation}
G_{a,b}(q,z) = \langle q,a | \frac{1}{z-(\hat{h}-E_0)} | q,b\rangle
\; .
\label{Greenfunction}
\end{equation}
For the structure factor in leading order, Eq.~(\ref{qstatesleadingorder})
requires 
the four Green's functions $G_{a,b}(q,\omega+{\rm i}0^+)$
for $a,b=1,L-1$. The second order requires the Green's function
for $a,b=1,2,L-2,L-1$. 
This cluster principle generalizes to higher orders, i.e., in $n$th
order we have to calculate a $(2n) \times (2n)$~matrix of Green's functions
for a potential which links $2n$ neighboring sites.

The Green's function of a particle in a 
potential of finite range is readily calculated~\cite{Ec83}.
We start from $\hat{h}=\hat{t}+\hat{V}$,
where $\hat{t}$ describes the
free particle motion over the ring with dispersion relation
$\varepsilon_q(k)$; up to second order,
we have
$\varepsilon_q^{(2)}(k)=-2t[\cos(k)+2\cos(k-q)]+13(t^2/U)-4(t^2/U)
[\cos(2k)+\cos(2k-2q)]$ with $0\leq k<2\pi$.
In the thermodynamic limit, the free Green's function is readily calculated,
\begin{equation}
g_{a,b}(q,z) = \langle q,a | \frac{1}{z-\hat{t}} | q,b\rangle
=\int_{-\pi}^{\pi} \frac{{\rm d}k}{2\pi}
\frac{
e^{ {\rm i} k(a-b)}}{z-\varepsilon_q(k)}  \;, 
\end{equation}
where we use the fact that the free states are plane waves.
We calculate the free Green's functions for $z=\omega\pm {\rm i}0^+$
with the help
of the residue theorem. Therefore, their real and imaginary parts
are available with high accuracy for all real frequencies~$\omega>0$.

With the help of the operator identity
\begin{equation}
\frac{1}{z-\hat{t}-\hat{V}}=\frac{1}{z-\hat{t}}+
\frac{1}{z-\hat{t}}\hat{V}\frac{1}{z-\hat{t}-\hat{V}}
\end{equation}
we derive the Green's function~(\ref{Greenfunction})
from the equation
\begin{equation}
G_{a,b}(q,z) = g_{a,b}(q,z)+ \sum_{l,m}g_{a,l}(q,z)V_{l,m}G_{m,b}(q,z) \; .
\end{equation}
This matrix equation has the formal solution 
\begin{equation}
\underline{\underline{G}}(q,z) 
= \left( 
\underline{\underline{1}} -\underline{\underline{g}}(q,z)
\underline{\underline{V}}  
\right)^{-1}
\underline{\underline{g}}(q,z) \; .
\label{matrixproblem}
\end{equation}
In $n$th-order perturbation theory, the potential 
and the required Green's functions have the same 
range~$2n$ on the lattice. Therefore,
the matrix problem in~(\ref{matrixproblem})
reduces to the inversion of a $2n\times 2n$-matrix for fixed $(q,z=\omega\pm 
{\rm i}0^+)$.

The Green's function calculation provides
higher-order corrections to $S^{[1]}(q,\omega)$. 
In Sect.~\ref{sec:results}
we show results for the dynamical structure factor
to fifth order in the $(t/U)$-expansion 
in the region around $\omega=U$.

The Green's function calculation does
{\sl not\/} cover the higher Hubbard sub-bands.
The first contribution to the structure factor
$S(q,\omega)$ beyond $\omega\approx U$ occurs 
around $\omega=3U$ with intensity $x^4$.
In the regions where strong-coupling perturbation theory 
is reliable, $x\lesssim 0.15$, the secondary
bands contribute only a few percent of the total weight.
When we include the term to order $x^4$ in the 
frequency-integrated structure factor, the sum-rule
for the structure factor is fulfilled, i.e., we reproduce
the terms $D(r)$ of the ground-state density-density correlation
function for $r=1,2,3$~\cite{DZ06} up to and including $x^{4}$.

\section{Variational Cluster Approach}
\label{sec:VCA}
The basic idea of the VCA is to approximate the
self-energy $\Sigma$ of a strongly correlated, 
physical system $\hat H$ by the self-energy
of an exactly solvable reference system $\hat
H'$~\cite{PAD03}. 
Both the physical and the reference system share the same
interaction but differ in their single-particle terms. 
The optimal self-energy is 
determined self-consistently from a stationary condition on the 
grand-canonical potential $\Omega$,
\begin{equation}
\frac{\delta \Omega}{\delta \Sigma}=0\;.
\label{var_gp}
\end{equation}
To evaluate this expression,
the self-energy is parameterized by the single-particle parameters of the
reference system.
In fact, this idea is quite general and allows to
unify (cluster)-Dynamical Mean Field Theory and VCA within the same theoretical 
framework depending on the choice of the reference system~\cite{Pot03a,Pot03b}.
In the case of the VCA, the reference system is chosen to be a cluster
decomposition of the physical system with modified single-particle parameters. 
Furthermore, the reference system is selected such 
that it can be solved exactly.  In principle, any many-body cluster 
solver at hand can be used which provides the dynamic 
single particle Green's function. Here, we use 
Lanczos exact diagonalization~\cite{ZEAH02,KAL10a}. 

Originally, VCA was introduced for fermionic systems~\cite{PAD03}. 
For correlated lattice bosons, it first has been in use to investigate 
the normal, Mott insulating phase~\cite{KD06}. 
In Refs.~\cite{KAL11,AKL11} VCA has been extended 
to the superfluid phase. 
This extension adopts the Nambu notation 
and is applicable to a large class of lattice models 
that exhibit a condensed phase. Since VCA is in the end a 
form of a cluster mean-field approach, it can obviously not comprise 
fluctuations at length scales larger than the cluster size. 
This means that in the case of power-law decaying correlations 
as present here, these are 
spuriously replaced by long-range order in the VCA. This is a common 
issue of all mean-field like approaches. Despite this drawback, VCA 
still provides reliable results for many observables such 
as the ground-state energy, the sound velocity of the phonon excitations, 
and the single-particle spectral function. 

Explicitly, the grand potential for bosonic systems with normal and 
the superfluid components is given by~\cite{KAL11,AKL11}
\begin{eqnarray}
2\Omega &=& 2\Omega^{\prime} - \Tr\ln (-G)  + \Tr\ln (-G')  
- \Tr (\underline{\underline{ t}} - \underline{\underline{ t}}')
\nonumber \\
&& +\langle \hat{A}\rangle^{\dagger}  [G\zero]^{-1}  
\langle \hat A \rangle 
- \langle \hat {A'}\rangle^{\dagger} [G\zero^{\prime}]^{-1}
\langle\hat A'\rangle \;,
\label{eq:omVCA}
\end{eqnarray}
where $G$ is the interacting Green's function, 
$\underline{\underline{t}}$ is the single-particle
Hamiltonian matrix, $\langle \hat{A}\rangle$ denotes 
expectation values of the Nambu boson operators 
consisting of both creation and annihilation operators, 
and the subscript `(0)' indicates that the Green's function 
is evaluated at zero wavevector and zero frequency. 
In~\eqref{eq:omVCA}, the prime marks  
again reference system quantities. 
The first line of~\eqref{eq:omVCA} is identical to the expression
in the normal phase~\cite{KD06} apart from the fact that the Green's
functions are considered to be in Nambu space and thus contain anomalous
parts which also account for the factor $1/2$.
The second line takes care of the condensation of bosons,
which in one dimension is an artifact of the dynamical and self-consistent 
mean-field treatment, as discussed above. 

To obtain the results presented below, 
we always use the chemical potential of the
reference system and a field which breaks the $U(1)$ symmetry on the level of
the reference system as variational parameters. In the Mott phase we also
determine the intercluster hopping and the boundary energies of the reference
system self-consistently~\cite{KAL10b}. Having found
the stationary point of the grand potential $\Omega$ with respect to the
variational parameters, we evaluate the dynamical single particle Green's
function $G(k,\omega)$ of the physical
system~\cite{KAL11,AKL11}. From that we calculate the single-particle
spectral function $A(k,\omega) = -\im G(k,\omega)/\pi$.
The static density-density correlation functions can be obtained from
the Fourier transform of the momentum distribution function, 
as specified in Ref.~[\onlinecite{KAL10b}].

\section{DMRG Approach}
\label{sec:DMRG}
The DMRG allows us to calculate static, dynamic and spectral  
properties of the 1D Bose--Hubbard model with high precision 
for fairly large system sizes. The main obstacle is related to the fact that,
in principle, each lattice site can be occupied by infinitely many 
bosonic particles. Therefore, one has to introduce a cutoff 
$n_b$, the maximum number of bosons per site taken into account. 
The DMRG results are nonetheless unbiased 
and numerically exact, if the dependence on $n_b$ can be proven to be
negligible and a careful finite-size extrapolation 
to the thermodynamic limit ($L\to\infty$) has been performed. 

Within the ground-state DMRG technique~\cite{Wh92,Wh93} 
the energy functional
\begin{eqnarray}
 E(\psi)=\frac{\langle\psi|\hat H|\psi\rangle}{\langle\psi|\psi\rangle}
\label{ene_functional}
\end{eqnarray}
is minimized in a variational subspace in order to find the ground-state
wave function $|\psi_0\rangle$ and energy $E_0=E(\psi_0)$ whereby the
DMRG energy error is proportional to the weight of the density-matrix
eigenstates discarded in the renormalization process. Increasing 
the number $m$ of density-matrix eigenstates kept, the 
discarded weight can be reduced systematically. 
Practically, the ground-state DMRG procedure mostly consists 
of two steps. During the infinite-system algorithm the system size 
is enlarged by two sites at each step and this operation has to be 
continued until the whole system reaches the desired system size $L$. 
Subsequently, a finite-system algorithm is used, where several sweeps 
through a lattice of fixed size $L$ are performed.
Thereby, the lattice is divided in two blocks with 
$\ell$ respectively $L-\ell$ sites where $1 \leq \ell\leq L-1$.
This sweeping improves the quality 
of the results obtained in the infinite-system algorithm. 
We note that this procedure is perfectly suited to compute the von Neumann 
entanglement entropy on-the-fly in the finite-system algorithm. From 
this quantity, the KT transition point 
of the Bose--Hubbard model can be determined accurately. 

The DMRG procedure can also be used to minimize the
following functional~\cite{Je02b,JF07}: 
\begin{eqnarray}
 W_{A,\eta}(\omega,\psi)
  &=&\langle\psi|(E_0+\omega-{\hat H})^2+\eta^2|\psi\rangle
  \nonumber \\
  && +\eta\langle A|\psi\rangle+\eta\langle \psi|A\rangle\;.
\end{eqnarray}
Here, $\hat{H}$ is the (time-independent) Hamilton operator and 
$\hat{A}$ denotes the quantum operator of the physical 
quantity to be analyzed; $\hat{A}^\dagger$ is its 
Hermitian conjugate and $|A\rangle=\hat{A}|\psi_0\rangle$.   
Once this minimization has been carried out, 
the dynamical correlation function 
\begin{eqnarray}
 G_{\hat{A}}(\omega+{\rm i}\eta)
  =-\frac{1}{\pi}\langle\psi_0|\hat{A}^{\dagger}
     \frac{1}{E_0+\omega+{\rm i}\eta-\hat{H}}\hat{A}|\psi_0\rangle,
\nonumber\\
\end{eqnarray} 
can be evaluated. Here, the small real number $\eta$ 
shifts the poles of the correlation function in the complex plane,
i.e., $\eta$ leads to a Lorentzian broadening of the peaks of the 
corresponding spectral function given in Lehmann representation as
\begin{eqnarray} 
I_{\hat{A}}(\omega+{\rm i}\eta)&=& {\rm Im} G_{\hat{A}}(\omega+{\rm i}\eta)\\
&=&
\frac{1}{\pi}\sum_n|\langle\psi_n|\hat A|\psi_0\rangle|^2\frac{\eta}{
(E_n-E_0-\omega)^2+\eta^2}\;.\nonumber
\end{eqnarray}
Within this so-called dynamical DMRG (DDMRG) technique, 
the sweeps in the finite-system algorithm are repeated until 
both functionals, $E(\psi)$ and $W_{A,\eta}(\omega, \psi)$, 
take their minimal values.

Investigating the Bose--Hubbard model by DMRG, 
we keep up to $n_b=5$ bosonic particles per site. 
Furthermore, we use $m=2000$ density-matrix
eigenstates in the DMRG runs for the ground-state 
expectation values. Then, the discarded weight 
is typically smaller than $10^{-10}$. In the DDMRG 
calculations we keep $m=500$ states to determine
the ground state during the first five DMRG sweeps, 
and afterwards use $m=300$ states for the calculation 
of the dynamical properties. 

\section{Results and Discussion}
\label{sec:results}

\subsection{von Neumann entanglement entropy}
\label{subsec:EE}

Previously, the KT transition point between the superfluid and insulating 
phases has been determined from the Luttinger parameter $K_b$~\cite{Ha81,MFZ08},
which can be extracted from the density-density correlation function
by DMRG, yielding  $t_c=0.305\pm0.001$ ($t_c=0.180\pm0.001$)
for $\rho=1$ ($\rho=2$)~\cite{EFG11}. Although $K_b(t<t_c)$
is not defined in the MI, $K_b(L)$ is finite and continuous over the KT
transition because the Mott gap is exponentially small. Therefore,
it can be used within a DMRG finite-size extrapolation procedure.

The quantum phase transition should  
become manifest in the system's entanglement properties 
as well~\cite{WBS06,RLSL12}. 
An important measure to quantify the entanglement of two subsets
of an interacting quantum system is the von Neumann entanglement entropy,
which shows a logarithmic scaling for critical systems~\cite{Ca04}.
To determine the critical point between a Tomonaga--Luttinger liquid 
and gapped (or ordered) phases for more subtle situations, 
e.g. for frustrated spin models, spinless fermion models with 
nearest-neighbor interaction or fermion-boson transport models, 
the use of the entanglement entropy difference has been demonstrated 
to be advantageous~\cite{Ni11,EBHEFS11,FEWB12}. 

\begin{figure}[tbp]
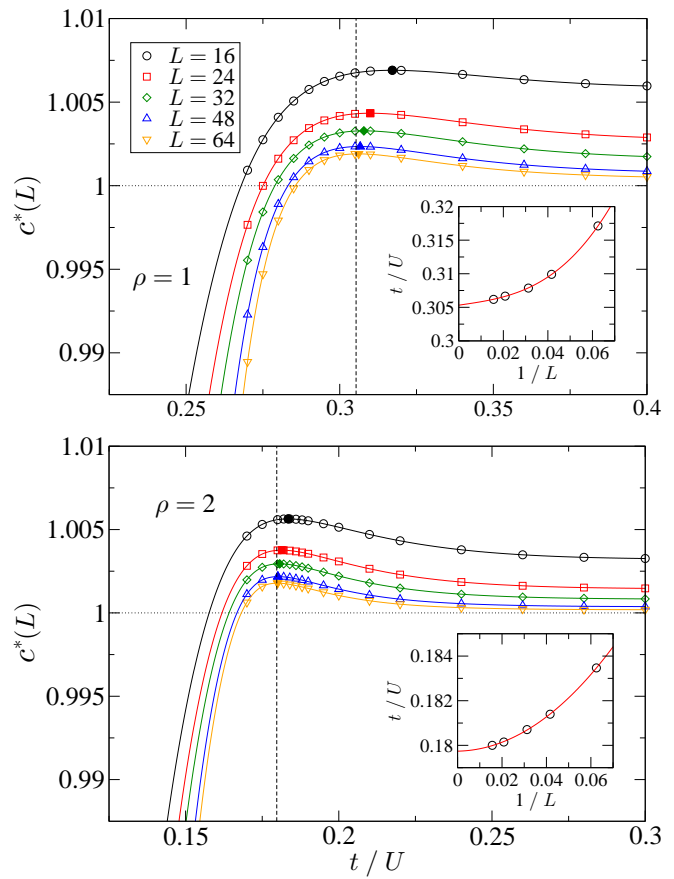

 \begin{center}
  \includegraphics[width=\columnwidth]{fig2_central_charge_rho1.eps}\\[6pt]
  \includegraphics[width=\columnwidth]{fig2_central_charge_rho2.eps}
 \end{center}
 \caption{(Color online) Entanglement entropy difference, $c^*$ from
Eq.~\protect\eqref{eq-cstar}, for the 1D Bose--Hubbard model 
with $\rho=1$ (upper panel) and $\rho=2$ (lower panel).
Data obtained by DMRG for lattices up to $L=64$ with PBC. 
The closed symbols indicate the maximum value for each system size. 
An extrapolation of the $t/U$ values at these maxima to the thermodynamic 
limit provides the Kosterlitz--Thouless
transition point, see insets; here, the lines correspond to a polynomial fit. 
The vertical dashed lines in the main panels mark the Kosterlitz--Thouless
transition point.
\label{KTtransition}}
\end{figure}

For a block of length $\ell$ in a periodic system of the system size $L$,
the von Neumann entropy, $S_L(\ell)$, is given by
$S_L(\ell)=-{\rm Tr}_\ell (\rho_\ell\ln\rho_\ell)$, 
with the reduced density matrix $\rho_\ell={\rm Tr}_{L-\ell}(\rho)$.
One finds for PBC~\cite{CC04}, 
\begin{eqnarray}
 S_L(\ell)=\frac{c}{3}\ln\left[
             \frac{L}{\pi}\sin\left(\frac{\pi\ell}{L}\right)
           \right]
           +s_1\;,
\label{von_neumann_entropy}
\end{eqnarray}
where $c$ is the central charge. 
When one evaluates the entropy
difference $S_L(L/2)-S_{L/2}(L/4)$
using DMRG with open boundary conditions (OBC)~\cite{LK08}, it 
includes the effect of the non-universal constant $s_1$. 
Therefore, the values for $t_c$
cannot be extrapolated systematically.
Here, we follow the alternative scheme proposed by Nishimoto~\cite{Ni11}.
We subtract $S_L(L/2)$ from $S_L(L/2-1)$ to obtain 
\begin{eqnarray}
 c^*(L)\equiv
\frac{3\left[S_L(L/2-1)-S_L(L/2)\right]}{\ln\left[\cos(\pi/L)\right]}\;.
\label{eq-cstar}
\end{eqnarray}
As $L\to\infty$, in the SF regime, the quantity $c^*(L)$ scales to the central 
charge $c=1$~\cite{CC04,LSCA06}.  

Fig.~\ref{KTtransition} displays $c^*(L)$ for
the 1D Bose--Hubbard model. Advantageously, we can use periodic boundary
conditions for the calculation of this quantity.
As shown in the insets, the position of the maximum in $c^*$ can be 
reliably extrapolated to the thermodynamic limit. 
In this way we get the cone point of the Mott lobes 
$t_c=0.305(3)$ for $\rho=1$ and $t_c=0.179(7)$ for $\rho=2$ (in  units of $U$),
in excellent agreement with the previous estimates 
from the OBC finite-size scaling of $K_b$~\cite{EFG11}.

\subsection{Ground-state properties}
\label{subsec:gsp}

\subsubsection{Ground-state energy}
\label{subsubsec:E0}

The ground-state energy $E_0$ of the 1D Bose--Hubbard model
has been determined analytically in the weak- and strong-coupling cases. 
For weak interactions, the Bogoliubov result was given in 
Eq.~\eqref{fullformula}, with the small-$U$ expansion given by Eq.~\eqref{sue}.
For strong interactions, an expansion up to 14th order in $x=t/U$
was obtained by Damski and Zakrzewski~\cite{DZ06}:
\begin{eqnarray}
 \frac{E_0^{[14]}}{4UL}&=& -x^2+x^4+\frac{68}{9}x^6
                 -\frac{1267}{81}x^8+\frac{44171}{1458}x^{10} \nonumber\\[2pt]
         && -\frac{4902596}{6561}x^{12}-\frac{8020902135607}{2645395200}x^{14}
\\[2pt]
&&   +{\cal O}(x^{16}) \nonumber \;.
\label{expansion-gs}
\end{eqnarray} 

\begin{figure}[b]
 \begin{center}
  \includegraphics[clip,width=\columnwidth]{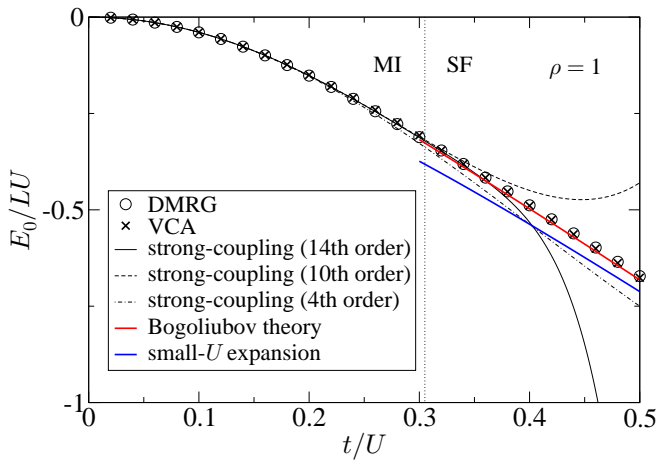}
 \end{center}
\caption{(Color online) 
Ground-state energy,  $E_0/(LU)$, as a function of 
interaction strength $t/U$ for $\rho=1$. 
Weak-coupling and strong-coupling results are compared with
the $L\to\infty$ extrapolated DMRG data obtained 
from chains up to $L=128$ with OBC. For the VCA calculations 
(crosses) a cluster with $L_c=12$ ($L_c=4$) sites is used in 
the MI (SF) phase.\label{fig:gs}}
\end{figure}

Fig.~\ref{fig:gs} compares these perturbative results with our 
VCA and DMRG data. The VCA reproduces the 
DMRG results almost perfectly for all interaction strengths $t/U\leq 0.5$.
The strong-coupling series expansion is also in accordance 
with the numerical exact data, surprisingly even beyond the KT 
transition point, i.e., for $t/U\lesssim 0.4$. 
Note that in Ref.~[\onlinecite{DZ06}] the ground-state 
energy~\eqref{expansion-gs} was compared with numerical data 
obtained for a system with $L=40$ sites only. Hence, in their figure,
the deviation starts at about $t\approx 0.2 U$. Clearly, the quality of the
strong-coupling approximation improves as higher-order corrections
are taken into account, cf.\ the 4th-, 10th-, and 14th-order results. 
Fig.~\ref{fig:gs} also shows
the range of validity of the corresponding weak-coupling approaches.
Surprisingly, the Bogoliubov result is applicable up to the 
Mott transition point.

\subsubsection{Boson correlation function}
\label{subsubsec:bcf}

In order to characterize the correlations in the ground-state
of the interacting Bose gas described by the Bose--Hubbard model,
it is instructive to look at the distance dependence of the expectation
values $\langle \hat{b}^\dagger_j\hat{b}_\ell^{\phantom{\dagger}}\rangle$,
which, with appropriate normalization, constitute the matrix 
elements of the one-particle density matrix~\cite{KSDZ04}.

\begin{figure}[hb]
\begin{center}
\mbox{}\\[6pt]
  \includegraphics[clip,width=\columnwidth]{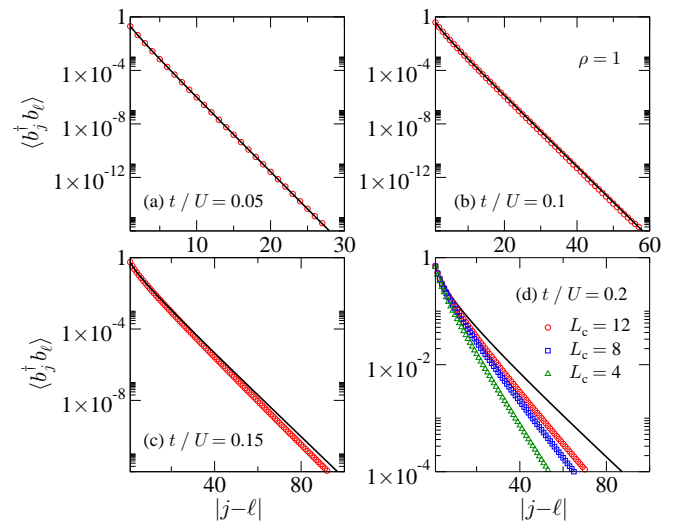}\\
\mbox{}
 \end{center}
 \caption{(Color online)
Decay of bosonic correlations in the 1D Bose--Hubbard model
within the first Mott lobe  ($\rho=1$) for decreasing interaction strengths 
$x=0.05$ (a), 0.1 (b), 0.15 (c), and 0.2 (d). 
DMRG results are obtained for a chain with $L=128$ sites and OBC.
To minimize the boundary effects we place $j$ and $\ell$ 
symmetrically around the center of the system. The VCA data  
were calculated using clusters with $L_c=4$ (green triangles), 
8 (blue squares), and 12 (red circles). \label{corr_func}}
\end{figure}

In the gapless SF state, the boson single-particle correlation function
 \begin{equation}
\langle \hat{b}^\dagger_j\hat{b}_\ell^{\phantom{\dagger}}\rangle
  \sim |j-\ell|^{-K_b/2}
\label{bcf}
\end{equation}
shows a power-law decay with an exponent determined by the Tomonaga--Luttinger 
parameter $K_b$~\cite{Gi04}. 

In the insulating (gapped) MI, the bosonic correlations
decay exponentially (at large distances), which is 
demonstrated by the semi-logarithmic representation in Fig.~\ref{corr_func}. 
At very strong couplings, the excitation gap is large
and therefore can be obtained very accurately within VCA.
As $x$ becomes larger, i.e., $U$ becoming smaller at fixed $t$,
the correlations are significant over many lattice sites. In this regime, 
we find noticeable deviations of the VCA results if $L_c$ is too small, 
see panel~(d).

\subsubsection{Momentum distribution function}
\label{subsec:MDF}

\begin{figure}[ht]
 \begin{center}
  \includegraphics[clip,width=0.47\textwidth]{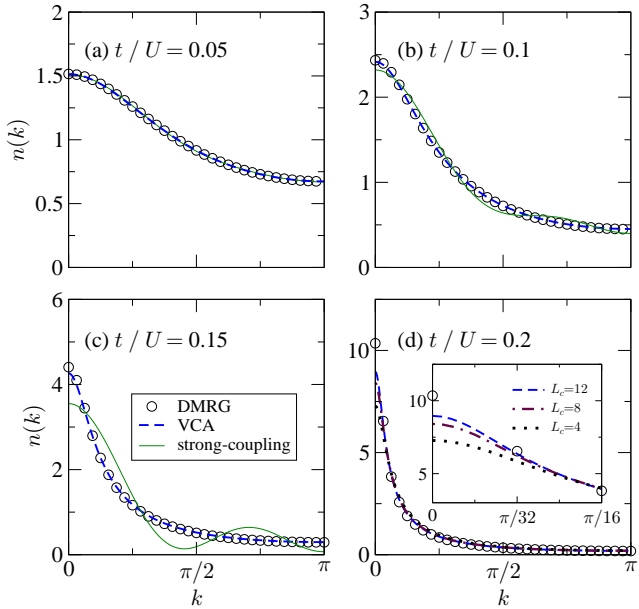}
 \end{center}
 \caption{(Color online) 
 Momentum distribution function $n(k)$ within the first Mott lobe 
 from DMRG with $L=64$ and PBC (symbols), VCA (dashed lines), and
 third-order strong-coupling expansion~\eqref{3rd-nk} (solid lines).
 The inset in panel (d) shows the dependence of the VCA results 
on the cluster size $L_c$ for $k\gtrsim 0$ in comparison with the 
DMRG data.\label{nk_MI}}
\end{figure}

The Fourier-transformed single-particle density
matrix gives the momentum distribution function 
\begin{equation}
 n(k)=\frac{1}{L}\sum_{j,\ell=1}^{L} e^{{\rm i}k(j-\ell)}
       \langle
        \hat{b}_j^\dagger\hat{b}_{\ell}^{\phantom{\dagger}}
       \rangle\;.
\end{equation}
To third order in $x=t/U$, strong-coupling theory predicts
for the first Mott lobe~\cite{DZ06,FKKKT09}: 
\begin{eqnarray}
n^{[3]}(k)&=&1+(8x-16x^3)\cos(k) \nonumber 
\\
&&+36 x^2 \cos(2k)+176 x^3\cos(3k)\;.\label{3rd-nk}
\end{eqnarray}
In Fig.~\ref{nk_MI} we compare the strong-coupling expansion~(\ref{3rd-nk}) 
with the DMRG and VCA numerics. 

While for $t/U=0.05$, where the momentum distribution 
is rather flat indicating weak site-to-site correlations, 
all methods essentially agree [see panel (a)], 
small deviations between analytical and numerical approaches appear
for $t/U\gtrsim 0.1$ [panel (b)]. The oscillations emerging for $x\sim 0.15 $
in the third-order strong-coupling theory are clearly an artifact.
The VCA reproduces 
the density distribution $n(k)$ very well. 
However, it fails quantitatively for $k\to 0$ and $x=0.2$ 
if the cluster used is not large enough, see the inset in panel (d).
  
When we approach the MI-SF KT transition point by raising $x$,
$n(k=0)$ will rapidly increase with system size. In 1D, of course, 
$n(k=0)$ will not attain a macroscopic value in the thermodynamic limit
because no true condensate develops~\cite{MC03}. Instead, we have 
from~Eq.~\eqref{bcf}
\begin{equation}
n(|k|\to 0)\sim |k|^{-\nu}\; , \; \nu=1-K_b/2<1\;.
\end{equation}
Thus far, it is difficult to reproduce this algebraic divergence
in the SF phase; see, however, Ref.~[\onlinecite{BEEFHS11}],
where $\nu$ was determined by DMRG.

\subsection{Dynamical quantities}
\label{subsec:DQ}

\subsubsection{Photoemission spectra and density of states}
\label{subsubsec:PES}

The single-particle excitations associated with the injection
and emission of a boson with wave vector $k$ and frequency $\omega$,
are described by the spectral functions $A^+(k,\omega)$ and $A^-(k,\omega)$,
see Eqs.~\eqref{A_plus} and~\eqref{A_minus}, respectively. These quantities
can be evaluated by VCA~\cite{KAL11} and DDMRG~\cite{Je02b,JF07}. 
For the Bose--Hubbard model  the following sum-rules hold 
[cf.\ Eqs.~\eqref{sr1}, \eqref{sr2}]:
\begin{eqnarray}
\int_{-\infty}^{\infty} {\rm d}\omega [A^+(k,\omega)-A^-(k,\omega)]&=&1\;,\\
\int_{-\infty}^{\infty} {\rm d}\omega A^-(k,\omega)&=&n(k)\;.
\end{eqnarray} 
Summing over momenta~$k$, 
the density of states $N(\omega)$ follows as
\begin{eqnarray}
 N(\omega)=A^{+}(\omega)-A^{-}(\omega)\;,
\end{eqnarray}
where $A^{\pm}(\omega)=\sum_k A^{\pm}(k,\omega)/L$. 
Within the DDMRG framework, however, it is much more appropriate 
to calculate $N(\omega)$ directly, instead of performing the $k$-summation
of $A^{\pm}(k,\omega)$.

First, we discuss the spectral function,
$A(k,\omega)=A^+(k,\omega)+A^{-}(k,\omega)$, and the density of states,
$N(\omega)$, in the MI regime. The DDMRG spectra for fixed $k$ 
consist of two Lorentzians of width $\eta=0.04U$, which
is the broadening introduced in the DDMRG procedure, cf.\ Sect.~\ref{sec:DMRG}.

\begin{figure}[t]
 \begin{center}
  \includegraphics[clip, width=0.47\textwidth]{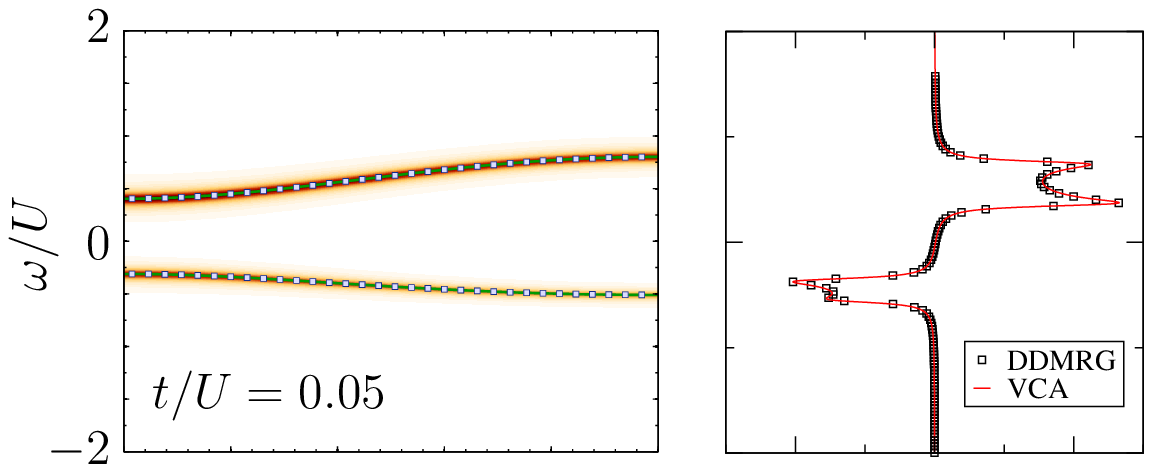}\\
  \includegraphics[clip, width=0.47\textwidth]{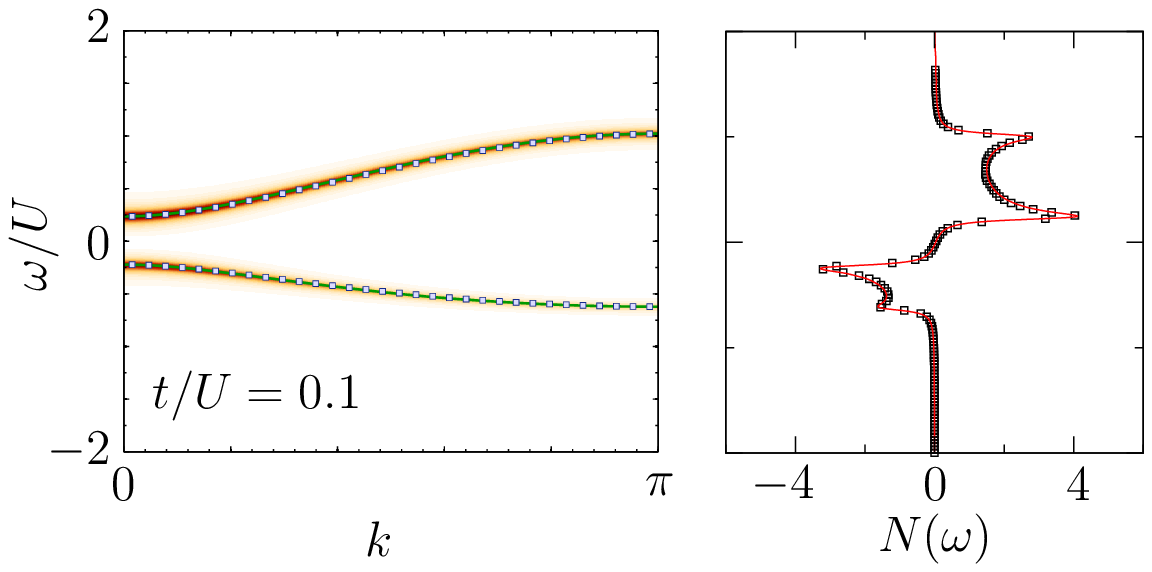}
 \end{center}
  \caption
  {(Color online) Single-particle spectral function  
$A(k,\omega)$ in the $k$-$\omega$ plane 
(left panels) and density of states $N(\omega)$ (right panels)
   for the 1D Bose--Hubbard model with $\rho=1$ and 
$t/U=0.05$ (upper panels), $t/U=0.1$ (lower panels).
We compare DDMRG data for a system with $L=64$ and OBC (squares)
with the results of VCA for $L_c=12$ and $\eta=0.03U$ 
(density plots) and strong-coupling expansions~(\ref{Ehole}) 
and (\ref{Eparticle}) (lines).\label{spec-strong}}
\end{figure}

Fig.~\ref{spec-strong} shows the quasiparticle dispersions (squares)
extracted from Lorentz fits to the maxima in $A^{\pm}(k,\omega)$. 
The quality of the fits suggests that the quasiparticle life-time 
is very large.
Because of the large Mott gap separating single-particle
and single-hole quasiparticle bands, the VCA can work with small cluster sizes
and the quasiparticle spectra are in perfect agreement 
with the DDMRG data. The same holds for the strong-coupling 
results. In fact, for large interactions, each site is singly occupied 
in the ground state. As a consequence, a hole or doubly occupied site
can move almost freely through the system. From this consideration, the
leading-order expression
for the quasiparticle dispersions results~\cite{EFG11}, 
see Eqs.~(\ref{Ehole}) and~(\ref{Eparticle}). 
We note in passing that the simple mean-field approach by 
van Oosten et al.~\cite{OSS01} fails to reproduce the 
quasiparticle dispersion already for $x=0.1$, see Ref.~[\onlinecite{EFG11}].

\begin{figure}[ht]
 \begin{center}
  \includegraphics[clip, width=0.48\textwidth]{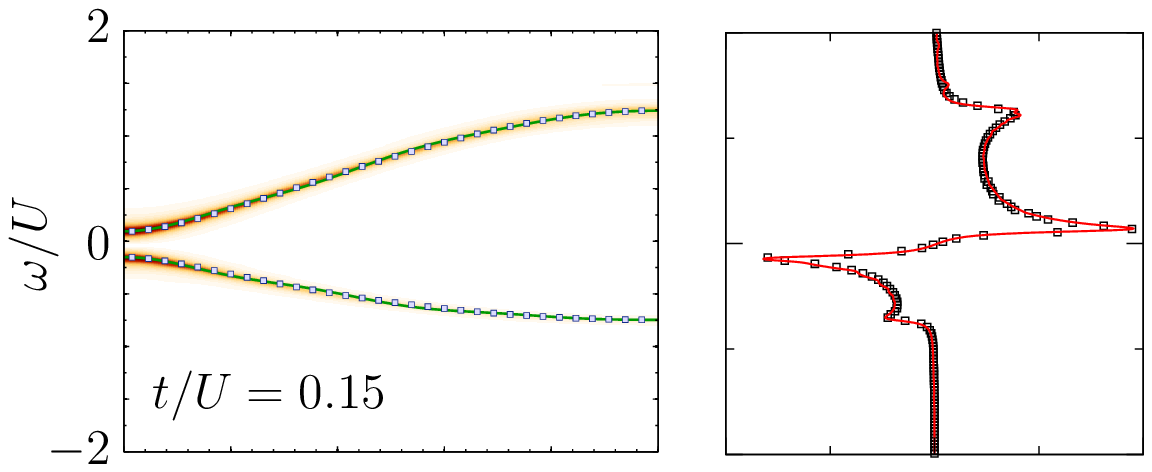}\\
  \includegraphics[clip, width=0.48\textwidth]{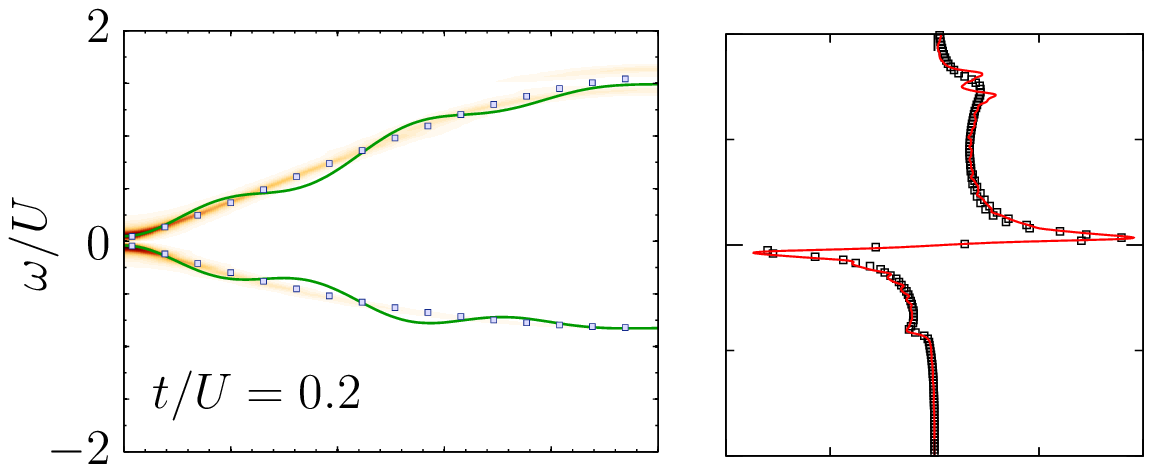}\\
  \includegraphics[clip, width=0.48\textwidth]{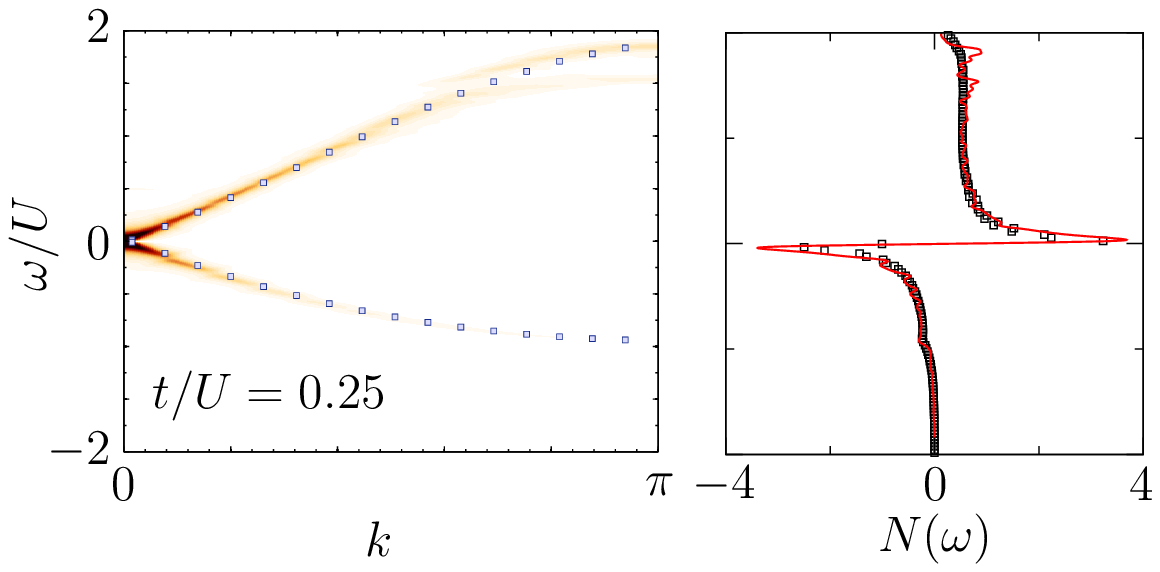}
 \end{center}
  \caption{(Color online) Spectral functions and density of states 
   for intermediate couplings $t/U=0.15$ (upper panels), 
   $t/U=0.2$ (middle panels) and $t/U=0.25$ (lower panels). 
Notations are the same as in Fig.~\ref{spec-strong}. 
\label{spec-intermediate}}
\end{figure}

As the on-site interaction further weakens, the Mott gap gradually closes;
the corresponding results are depicted in Fig.~\ref{spec-intermediate}.
Obviously, strong-coupling theory becomes imprecise at $x\approx 0.2$,
and completely fails at $x\gtrsim 0.25$. There also VCA shows some
artificial gap features near the Brillouin zone boundary, which do 
not show up in DDMRG.

\begin{figure}[t]
 \begin{center}
  \includegraphics[clip,width=0.48\textwidth]{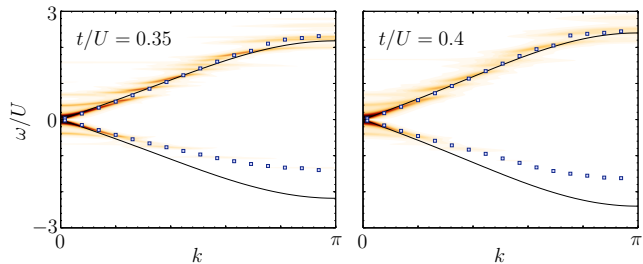}
 \end{center}
 \caption{(Color online)
 Spectral functions in the superfluid phase of the 1D Bose--Hubbard model 
 with $t/U=0.35$ (left panel), $t/U=0.4$ (right panel).
 Compared are DDMRG dispersions with VCA density plots for $L_c=4$
and the dispersion of the condensate excitations $E(k)$ from Bogoliubov 
theory (black lines, other symbols are the same as in 
Fig.~\ref{spec-strong});
see Eq.~\eqref{bE} with~\eqref{be}.\label{spec-SF}}
\end{figure}

In the superfluid phase, the elementary excitations concentrate around 
the region $(k=0,\omega=0)$, see Fig.~7 in Ref.~[\onlinecite{EFG11}], which
indicates the formation of a ``condensate''. In accordance with
Bogoliubov theory and field theory~\cite{Bo47,CCGOR11}, the low-energy, 
low-momentum excitations dominate the single-particle spectrum.
As can be seen from Fig.~\ref{spec-SF}, our spectral
function indeed exhibits a phonon mode
whose excitation energy --for a system in the thermodynamic limit-- 
is linear in $k$ and gapless at $k=0$. 
Yet, for finite-size systems a gap is present 
whose magnitude is inversely proportional to the system size. 
Our DDMRG data demonstrate that the gap almost vanishes already    
for a OBC system with 64~sites. A similar behavior has been observed 
in QMC calculations which employ the directed-loop method~\cite{PEH09}. 

Within the VCA, we find a 
larger gap as compared to DDMRG, due to the fact that we 
solve only four-site clusters exactly which are 
subsequently coupled perturbatively. In Ref.~[\onlinecite{KAL11}]
we showed that the gap at $k=0$ decreases with increasing cluster size, 
suggesting the correct behavior in the infinitely large cluster limit. 
Along the linear Goldstone modes, the spectral weight
obtained by means of VCA exhibits fringes and a series of minigaps.  
This behavior is most likely a result of the
cluster decomposition and subsequent periodization of the Green's
function~\cite{KAL10a}. However, it should be emphasized 
that the slopes of the phonon mode obtained by the two methods 
agree very well. 

A universal feature of systems with broken $U(1)$
symmetry is that, in addition to a gapless Goldstone mode, 
a gapped amplitude mode should be present. 
Whereas the Goldstone modes correspond to phase fluctuations, the
amplitude
modes arise from fluctuations in the magnitude of the order parameter. 
This behavior can be sketched by a Mexican hat potential for the order
parameter~\cite{Sa11}.
It has been argued in Ref.~[\onlinecite{Sa11}]
that the amplitude modes are sharp
excitations in the quasiparticle sense only for dimensions $d\geq 3$
for which they were detected experimentally~\cite{HofstetterPRL}.

For $d<3$ the decay of the amplitude modes into Goldstone modes
is very efficient and, thus, the weight observed in 
the susceptibilities can be 
redistributed over a large frequency range. This renders an observation 
of the amplitude modes difficult. 
In 2D it was demonstrated theoretically that the coupling 
to the amplitude modes can be improved by evaluating 
susceptibilities for the kinetic energies~\cite{PP12} 
or for operators that resemble the rotationally invariant 
structure of the Mexican hat potential~\cite{PAA11}. 
This should result in clearer signals for the amplitude modes 
in the respective response functions. 
Indeed, in setups with ultracold atoms, 
recent lattice modulation experiments, 
which couple directly to the amplitude modes, 
provide evidence for their existence in 2D~\cite{EFPCSGDKB12}.

In 1D where no true condensate exists, 
the spectral smearing of the amplitude modes is believed 
to be even more pronounced. 
Our numerical DMRG and VCA results for the spectral 
function and the dynamical structure factor, which both couple to 
the gapless Goldstone mode and the amplitude mode, 
give no indication for the latter. Therefore, we do not expect
that an amplitude mode will be observed
in Bragg spectroscopy experiments for bosons in one dimension.
It is quite remarkable that VCA reproduces the overall character 
of the single-particle spectrum consisting of Goldstone modes only, 
despite of the fact that, technically, a spurious condensate 
has to be introduced to treat the superfluid phase, see Sec.~\ref{sec:VCA}
for a detailed discussion. 

\begin{figure}[ht]
 \begin{center}
  \includegraphics[clip,width=0.95\columnwidth]{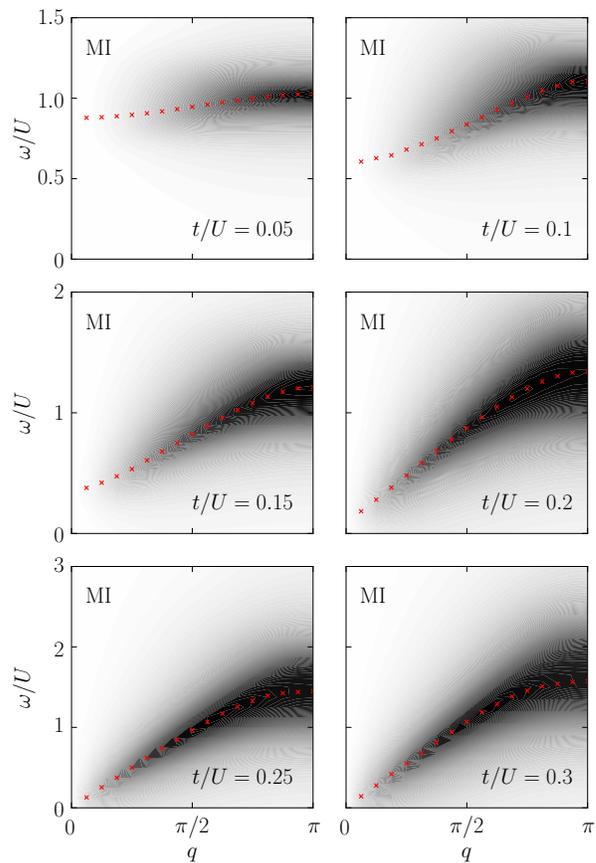}
 \end{center}
 \caption{(Color online) 
 Intensity of the dynamical structure factor $S(q,\omega)$ 
 in the MI phase of the 1D Bose--Hubbard model for different $t/U$ 
where $\rho=1$.
DDMRG data were obtained for an $L=32$ site system with PBC, using $\eta=0.5t$. 
Red crosses mark the positions of the maximum in each 
$q=2\pi m_q/L$-sector.
\label{SqwMI_density}}
\end{figure}

\subsubsection{Dynamic density-density correlations}
\label{subsubsec:dddc}

We now turn 
to the dynamical density-density response function. We carry
out large-scale DDMRG  calculations of the dynamic structure factor,
$S(q,\omega)$, and compare the results with the predictions of
strong-coupling theory~\eqref{Skomeagfromexacteigenstates} where appropriate.
In strong coupling, we show results in fourth-order approximation.
The agreement with the DDMRG data for $x=0.15$ improves noticeably when 
we calculate expectation values with
$|q\rangle\approx \sum_{n=1}^5 x^n |q^{[n]}\rangle$ from~(\ref{qstatesdef}),
i.e., we keep the states to fifth order in~$x$.

Since DDMRG provides $S(q,\omega)$ with a finite broadening $\eta$,
it turns out to be useful to convolve the strong-coupling result 
with the Lorentz distribution~\cite{NJ04},
\begin{eqnarray}
 S_{\eta}(q,\omega)=
  \int_{-\infty}^{\infty}{\rm d}\omega^{\prime} S(q,\omega^{\prime})
   \frac{\eta}{\pi\left[(\omega-\omega^{\prime})^2+\eta^2\right]}\;.
\end{eqnarray}

{\it Mott phase.}
Fig.~\ref{SqwMI_density} illustrates the change of the intensity distribution
of $S(q,\omega)$ in the $q$-$\omega$ plane as $x=t/U$ increases in the
MI regime. For small $x$, deep in the MI, the spectral weight is concentrated
around $\omega\sim U$ in the region $q>\pi/2$ (cf.\ the upper left panel). 
This meets the predictions of the strong-coupling theory~\cite{EFG12}. 
In this regime the structure factor is dominated by the primary band.  

\begin{figure}[t]
 \begin{center}
  \includegraphics[clip,width=\columnwidth]{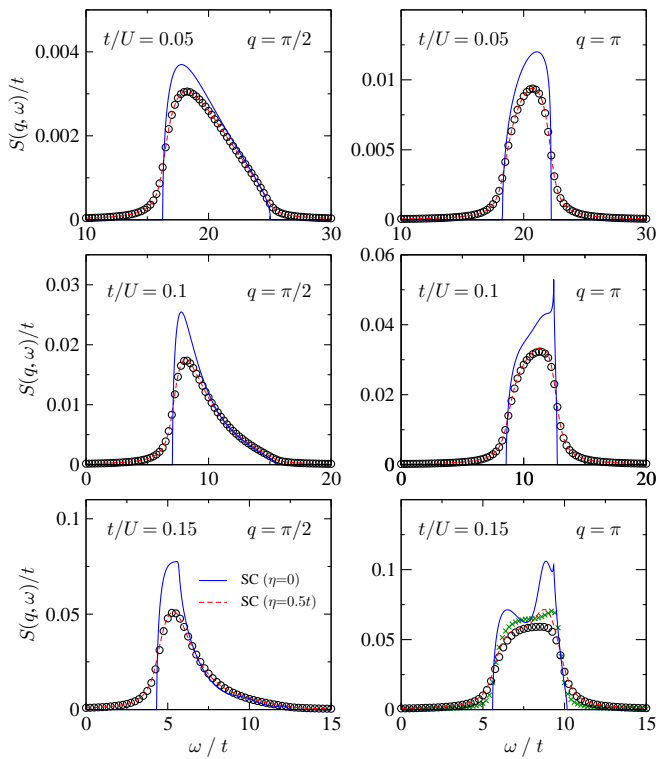}
 \end{center}
 \caption{(Color online) 
Frequency-scans of the dynamical structure factor in the MI state 
at fixed momenta $q=\pi/2$ (left panels),  $q=\pi$ (right panels).
DDMRG data (circles) were obtained for $L=64$, PBC, and $\eta=0.5t$;
at $t/U=0.15$ ($q=\pi$) also for $L=128$, PBC, and $\eta=0.2t$ 
(green crosses). Blue solid (red dashed) lines give the corresponding 
results of the strong-coupling theory with $\eta=0$
($\eta=0.5t$). 
Please note the different scales of the ordinates.\label{SqwMI} }
\end{figure}

When $x$ increases, the maximum of $S(q,\omega)$ acquires an appreciable   
dispersion; simultaneously the overall intensity of the density-density 
response strengthens, see the middle panels 
of Fig.~\ref{SqwMI_density} and also Fig.~\ref{SqwMI}. 
As the system approaches the MI-SF transition point, the excitation gap 
closes.  Concomitantly, we find a significant redistribution of 
the spectral weight to smaller $q$ values, see the lower panels
of Fig.~\ref{SqwMI_density}.

In Fig.~\ref{SqwMI} we show constant-moment
scans of $S(q,\omega)$ at $q=\pi/2$ and $q=\pi$.
For $x=0.05$ and $x=0.10$, the agreement between the broadened
strong-coupling results and the DDMRG data
for $S(q,\omega)$ is excellent. As $x$ becomes larger than $x\approx 0.10$,
the strong-coupling theory yields a double-peak structure 
in $S(\pi,\omega)$. When we
increase the lattice size and reduce~$\eta$, this feature also appears
in our DDMRG data for $t/U=0.15$. Therefore, this feature is not
an artifact of the strong-coupling approach
even though the strong-coupling expansion overestimates
the double-peak structure for $x=0.15$.
The strong-coupling expansion solves an effective single-particle problem
in a (finite-range, attractive) potential. Such a potential 
gives rise to a non-trivial spectrum (resonances and possibly bound states).
The energy levels of the effective single-particle problem lead to
non-trivial spectral signatures in the dynamical structure factor.

\begin{figure}[t]
 \begin{center}
  \includegraphics[clip,width=0.95\columnwidth]{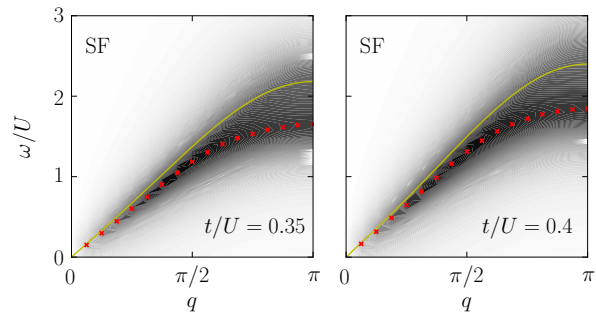}
 \end{center}
 \caption{(Color online) Intensity of the dynamical structure
  factor $S(q,\omega)$ in the superfluid phase of the Bose--Hubbard model
with $\rho=1$. Again we use $L=32$, PBC, and $\eta=0.5t$. 
  The yellow line gives the Bogoliubov result~\eqref{bE}.
\label{SkwSF} }
\end{figure}

\begin{figure}[t]
 \begin{center}
  \includegraphics[clip,width=0.95\columnwidth]{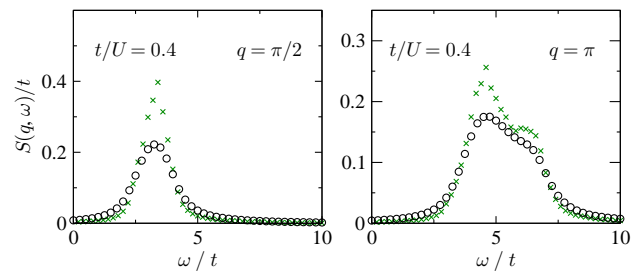}
 \end{center}
 \caption{(Color online)
 Frequency dependence of the dynamical structure factor 
$S(q,\omega)$ at $q=\pi/2$ and $q=\pi$. Only DDMRG data are shown.
Circles (crosses) mark the results for $L=64$, PBC, and $\eta=0.5t$ 
($L=128$, PBC, and $\eta=0.2t$).
\label{SqwSFt0.40qPi} }
\end{figure}

{\it Superfluid phase.}
Figs.~\ref{SkwSF} and~\ref{SqwSFt0.40qPi} present the corresponding 
results for the dynamical structure factor
in the SF phase. At small momenta, Bogoliubov theory gives the correct
slope of the dispersion which we derive from the maximum of the
DDMRG data for $S(q,\omega)$. 
Note that the dispersion $E(q)$ in~(\ref{bE})
is identical to the predictions from field theory~\cite{Gi04}.
Bogoliubov's dispersion overestimates the DDMRG maxima
for larger momenta and higher energies, as observed experimentally
for a 3D setup~\cite{EGKPLPS09}.
As compared to the MI phase, 
the density-density response has higher intensity in the
SF state.
Interestingly, we also find a shoulder in $S(\pi,\omega)$, which may 
form a double peak as $L\to\infty$, $\eta\to 0$, see the
right-hand panel of Fig.~\ref{SqwSFt0.40qPi}. 
This high-energy double peak in the SF phase resembles the 
structure seen in the MI phase.
In our opinion, 
this rules out an interpretation of the second peak
as signature of a massive Higgs mode~\cite{HABB07}.

\section{Summary}
\label{sec:S}

The aim of this paper was twofold: (i)~to provide extensive
numerical (D)DMRG data for static and dynamical quantities of
the one-dimensional Bose--Hubbard model at integer filling, mostly for
$\rho=N/L=1$;
(ii) to compare the (D)DMRG results with the analytical strong coupling
perturbation theory and the numerically inexpensive VCA and thereby
explore their merits and limitations 
in the most demanding case of one dimension. 

We used the DMRG to calculate the central charge from which
we confirmed the critical values for the superfluid to Mott transition
for integer fillings $\rho=1$ and $\rho=2$.
 
The ground-state energy from DMRG compares favorably 
with results from VCA and from
perturbation theory. For static correlation functions
such as the single-particle density matrix and the momentum distribution,
the comparison between DMRG data and strong-coupling perturbation theory (VCA)
shows that the latter are reliable for $x=t/U
\lesssim 0.15$ ($x\lesssim 0.25$),
for doable implementations. 

We calculated dynamical quantities such as the 
single-particle spectral function
and the dynamical structure factor.
In the superfluid phase, the response at low energies
is dominated by the quasi-condensate, in agreement with predictions
from field theory and Bogoliubov theory. The latter provides the correct
result for the phonon mode despite the fact that it is based on
the incorrect assumption of a true condensate. For finite interactions
and at higher energies, the dynamical structure factor is broad and 
reflects the physics of the Mott insulator. The overall character 
of the single-particle spectrum and the sound velocity of the phonon modes
are reproduced by VCA for larger values of $x$.

The strong-coupling results for the dynamical structure factor 
helped us to interpret our numerical DDMRG data 
because the latter are spectrally broadened
for finite system sizes.
The two-particle correlation function in the Mott phase 
reflects the (scattering) states
of a doubly occupied site and a hole with a hard-core repulsion and
a (weak) longer-ranged attraction giving rise to a
double-peak structure in the dynamical
structure factor near the boundary of the Brillouin zone.

Our numerical work can be compared with experiments only after
the parabolic confinement potentials will have been taken into account.
Important as it is to confine the atoms to the optical lattice,
the confinement potential often is so strong that the density profile contains 
several Mott regions with different integer fillings and transition regions 
between them. In this case, the structure factor at low energies
describes the dynamical response of the `wedding-cake' density 
profile~\cite{batrouni_dynamic_2005,pupillo_bragg_2006}.

There are two other directions to extend our work.
The Mott gap in the Bose--Hubbard model resembles a band gap. 
Therefore, it is interesting to see how
bound states (`excitons') form in this gap 
in the presence of a nearest-neighbor attraction.
A second route to extend our work is the inclusion of a disorder 
potential~\cite{SchollwoeckZwerger,CCGOR11,knap_excitations_2010}
so that the smearing and closing of the Mott gap as a function
of the disorder can be studied.
Work in this direction is in progress.

\acknowledgments

The authors thank
I.\ Danshita, G.\ Hager, E.\ Jeckelmann, H.\ Monien, S.\ Nishimoto, and
W.\ Zwerger for valuable discussions.
Financial support by the Deutsche Forschungsgemeinschaft
through the SFB 652 and by the Austrian Science Fund (FWF) P24081-N16
 are gratefully acknowledged.

\end{document}